\documentclass{article}

\usepackage[aux]{rerunfilecheck}
\usepackage{amsmath}
\usepackage{amssymb}
\usepackage{mathtools}
\usepackage[numbers,sort&compress]{natbib}
\usepackage[autostyle]{csquotes}
\usepackage{xfrac}
\usepackage{float}
\usepackage{graphicx}
\usepackage[dvipsnames]{xcolor}
\usepackage{bbm} 
\usepackage{placeins} 
\usepackage{booktabs}
\usepackage[top=1in, bottom=1.25in, left=0.9in, right=0.9in]{geometry}
\usepackage{tikz}
\usetikzlibrary{decorations.pathreplacing}
\usetikzlibrary{angles}
\usetikzlibrary{shapes.geometric}   
\usepackage{rotating}    
\usepackage{siunitx}  
\usepackage{multirow}
\usetikzlibrary{tikzmark,fit}
\usepackage{subfigure}
\usepackage{enumerate}
\usepackage{hyperref}

\title{\bf Neutrino Oscillations as a Gravitational Wave Detector?}

\author{Dominik Hellmann$^1\footnote{dominik.hellmann@tu-dortmund.de}$\,, Sara Krieg$^1$\footnote{sara.krieg@tu-dortmund.de}\,, Heinrich P\"as$^1\footnote{heinrich.paes@tu-dortmund.de}$\,, Mustafa Tabet$^1$\footnote{mustafa.tabet@tu-dortmund.de}
\smallskip
\\
{\it $^1$ Fakult\"at f\"ur Physik,
Technische Universit\"at Dortmund,
Germany}
}

\begin{document}

\maketitle

\begin{abstract}
    Gravitational waves (GWs) can alter the neutrino propagation distance and thus affect neutrino oscillations.
    This can result in a complete disappearance of the oscillatory behavior that competes with other sources of neutrino decoherence.
    We develop a set of criteria that determines under which conditions neutrino oscillations are sensitive to this effect.
    We find that current or near future neutrino oscillation experiments are not sufficiently sensitive to 
    coherent GW signals but may probe the stochastic gravitational wave background if the energy resolution
    improves drastically by several orders of magnitude.
\end{abstract}

\section{Introduction}
\label{sec:intro}
Gravitational wave (GW) astronomy provides a new window for our exploration of the universe.
This includes the observation of previously inaccessible phenomena, such as the merger
of black holes~\cite{Eroshenko:2023bbe} and allows for the investigation and detection of neutron
star mergers~\cite{Casalderrey-Solana:2022rrn}, collisions between neutron stars and black holes~\cite{Rahvar:2023iup}, 
and potentially the explosion of supernovae in the future~\cite{Powell:2023bex}.
Moreover, GWs can be employed in the search for exotic objects such as cosmic strings~\cite{Eichhorn:2023gat}
or primordial black holes~\cite{LISACosmologyWorkingGroup:2023njw}, and thus provide clues to physics 
beyond the Standard Model (SM). In principle GWs in the sub $\si{\nano\hertz}$ 
range may even provide novel possibilities to constrain inflationary models~\cite{Guzzetti:2016mkm}.
Recent achievements like NANOGrav's detection of evidence for a GW
background using pulsar timing arrays (PTAs)~\cite{NANOGrav:2023gor} 
prove that we are at the beginning of an exciting new era of GW and 
multimessenger astronomy. Further efforts are underway to expand 
the capabilities of existing gravitational wave detectors such as LIGO 
and Virgo, while plans for next generation detectors like the Einstein Telescope~\cite{Branchesi:2023mws}
and LISA (Laser Interferometer Space Antenna)~\cite{Baker:2019nia} promise an even greater sensitivity 
and a broader range of observations in the near future.  

While much attention has been focused on the detection and
characterization of GWs using conventional methods, such 
as laser interferometry~\cite{LIGOScientific:2007fwp}, there is a growing recognition of the potential 
role that neutrinos could play in this area of research~\cite{Dvornikov:2019jkc,Koutsoumbas:2019fkn,Dvornikov:2019xok,Dvornikov:2020dst,Dvornikov:2021sac,Dvornikov:2022bwr,Lambiase:2022ucu,Lambiase:2023pxd,Domi:2024ypm,Mandal:2024}.
As has been already pointed out in references~\cite{Stodolsky:1978ks,Piriz:1996mu,Ahluwalia:1996ev,Ahluwalia:1996wb,Cardall:1996cd,Fornengo:1996ef,Lambiase:2005gt,Visinelli:2014xsa}, 
gravitational fields modify the time evolution phase of quantum particles. 
Since neutrinos are able to traverse vast distances through astrophysical environments without being deflected from their trajectory,
they could offer unique insights into the GW landscape. 
One promising research direction is to study how GWs affect
quantum interference phenomena like neutrino flavor oscillations.

So far, most works on this subject consider the influence of a stochastic GW background (SGWB)
as a source of quantum decoherence in neutrino oscillations, see for example references~\cite{Dvornikov:2019jkc,Koutsoumbas:2019fkn,Dvornikov:2019xok,Dvornikov:2020dst,Dvornikov:2021sac,Dvornikov:2022bwr,Lambiase:2022ucu,Lambiase:2023pxd}.
In this work we argue that not only stochastic perturbations of the neutrino evolution phase
can cause decoherence in the neutrino system, but that also GW signals from individual sources will 
at least in principle have a similar effect that may be searched for.
We discuss in detail the challenges and prospects of such an approach 
and investigate the importance of the interplay of GW induced decoherence with other sources
of decoherence, as described by references~\cite{Kayser:1981ye,Frampton:1982qi,Giunti:1991ca,Giunti:1993se,Giunti:1997sk,Kiers:1997pe,Akhmedov:2009rb,Akhmedov:2014ssa,Kersten:2015kio,Akhmedov:2017mcc,Naumov:2020yyv}.
These decoherence effects arise, for example, from the separation of neutrino wave 
packets, potentially macroscopic neutrino production regions and 
finite energy binning and result in a similar damping of the oscillation 
behavior of neutrino flavor transitions. Therefore it is important to account simultaneously for these effects 
when considering GWs from coherent sources or from the SGWB, a step that has not yet been undertaken in previous studies.


This paper is organized as follows. In section~\ref{sec:modProbability} 
we develop a simple yet robust heuristic model accounting 
for the effects of the curved spacetime induced by GWs
on neutrino propagation
by considering deviations from the static trajectory approximation.
A time average of the oscillation probability over the running time of a neutrino 
experiment is employed in order to account for a varying travel distance and to predict the expected neutrino oscillation pattern.
In section~\ref{sec:ObsCons} we develop several criteria that determine whether neutrino oscillations 
can be sensitive to GWs. We estimate the sensitivity of neutrino oscillation experiments to 
coherent GW signals by applying the developed criteria.
In section~\ref{sec:sgw} we study the influence of the SGWB on neutrino oscillations, starting
with the selection of neutrino sources by considering the previously developed criteria in section~\ref{ssec:sgwb_theory}.
The statistical analysis and toy experiment setup are presented in section~\ref{ssec:sgwb_stat_ana}, the 
corresponding results are discussed in section~\ref{ssec:SM_pulsar_SGWB}.
A summary and conclusions are provided in section~\ref{sec:conc}. 
Technical details regarding the calculation and the statistical method 
employed for the analyses can be found in the appendices.

\section{Modified Oscillation Probability in the Presence of Gravitational Waves}
\label{sec:modProbability}

To investigate how the propagation of neutrinos through a curved spacetime induced by GWs 
affects their flavor oscillations, we start with the standard neutrino oscillation
probability
\begin{align}
    \hat{P}_{ab}(E, L) = \sum_{j} \vert U_{a j} \vert^2 \vert U_{b j} \vert^2
    + 2 \sum_{j < k} \mathrm{Re}\left(U_{a j}^{\ast} U_{b j} U_{a k} U_{b k}^{\ast} \exp\left[-2 \pi i \frac{L}{L_{jk}^{\mathrm{osc}}} - \mathcal{D}_{jk}(E, L)\right] \right) \,, \label{eqn:prob0}
\end{align}
where \(\mathcal{D}_{jk}\) is a positive damping term describing the decoherence due to wave packet separation and \(L_{jk}^{\mathrm{osc}} = 4 \pi E / \Delta m_{jk}^2\) is the oscillation length. 
The matrix $U$ describes the neutrino mixing, e.g. for three neutrino flavors \(U\) corresponds to the Pontecorvo--Maki--Nakagawa--Sakata (PMNS) matrix, \(\Delta m_{jk}^2 = m_j^2 - m_k^2\) are the squared mass splittings of the neutrino mass eigenstates and \(E\) is the average energy of the neutrino system.
Depending on the shape of the neutrino wave packets the decoherence function \(\mathcal{D}_{jk}\) can differ.
A common assumption for the neutrino wave packets is that they are of Gaussian shape implying
\begin{align}
    \mathcal{D}_{jk}(E, L) = \left(\frac{L}{L_{jk}^{\mathrm{coh}}}\right)^2 \quad \text{with} \quad L_{jk}^{\mathrm{coh}} = 2\sqrt{2} \frac{\sigma_x}{\vert \Delta v_{jk} \vert} \quad \text{and} \quad \Delta v_{jk} = \frac{\Delta m_{jk}^2}{2 E^2} \,,
\end{align}
with the position space wave packet width \(\sigma_x\) that we take to be equal for all mass eigenstates for simplicity.

We now describe how GWs influence neutrino oscillations by modifying the length $L$ in the standard probability, cf. equation~\eqref{eqn:prob0}, 
and replacing it with the propagation distance of neutrinos in curved spacetime. For this, we consider the average geodesic \(\gamma\) 
that the neutrino wave packets follow which now are slightly perturbed due to the presence of the GW,
\begin{align}
    \gamma_{0}^{\mu}(\tau) &= \underbrace{\frac{X_{D}^{\mu} - X_{P}^{\mu}}{\tau_D - \tau_P}}_{u^{\mu}} (\tau - \tau_P) + X_{P}^{\mu} 
    \quad \longrightarrow \quad
    \gamma^{\mu}(\tau) = \gamma_{0}^{\mu}(\tau) + \Delta\gamma^{\mu}(\tau) \,,
\end{align}
where \(X_{P(D)}^{\mu}\) denote the coordinates of neutrino production (detection), \(\tau\) is the proper time parameterizing the neutrino geodesic and hence \(\tau_{P(D)}\) denote its corresponding value at production (detection), respectively.
To obtain the first order correction \(\Delta \gamma\) to the flat spacetime geodesic \(\gamma_0\), we solve the geodesic equation for the metric\footnote{In this work we use the west coast convention, i.e. \(\eta = \mathrm{diag}(+1,-1,-1,-1)\).} \(g_{\mu\nu} = \eta_{\mu\nu} + h_{\mu\nu}\) with the 
pertubation to flat space time $h_{\mu\nu}$, describing the GW, with boundary values\footnote{Especially we assume that this problem is well posed, i.e. that geodesics connecting those points are unique.} \(X_P^{\mu}\) and \(X_D^{\mu}\).
For the step by step calculation see appendix~\ref{app:Geodes}.

The spatial length $L$ that neutrino wave packets have traveled between production and detection coordinate time $t_{P}$ and $t_{D}$ is given by
\begin{align}
    L(t_P, t_D) &= \int_{\tau_P}^{\tau_D} \sqrt{- g_{jk}(\gamma(\tau)) \dot{\gamma}^{j}(\tau) \dot{\gamma}^{k}(\tau)} \;\mathrm{d}\tau \label{eqn:Length_metric}\\
    &\approx L_0 - \frac{u^0}{2} \int_{\tau_P}^{\tau_D} h_{\parallel}(\gamma_0(t)) \; \mathrm{d}\tau =: L_0 + \Delta L\,, \label{eqn:L1}
\end{align}
which is then substituted into the standard oscillation probability.
Here \(h_{\parallel}\) is the metric perturbation projected onto the zeroth order neutrino trajectory \(\gamma_0\).
Moreover, we use the ultrarelativistic approximation for the neutrino velocity and neglect terms of \(\mathcal{O}(h^2)\) in order to express the detection time as \(t_D \approx L_0 + t_P \equiv L_0 + t\).
For the details of the calculation see Appendix~\ref{app:ModPath}.
In order to evaluate the proper time intergration in equation~\eqref{eqn:L1}, we use the fact 
that any GW can be decomposed into plus and cross polarized plane waves, i.e.
\begin{align}
    h_{\mu\nu}(x) &= \sum_{r \in \{+, \times\}} \int \mathrm{d}^3\vec{k} \; \psi_r(\vec{k}) A_{\mu\nu}^r \cos(\omega t - \vec{k} \vec{x} + \phi^r(\vec{k})) \,,
\end{align}
where $A_{\mu\nu}^r$ are normalized polarization tensors, \(\psi_r(\vec{k})\) are real, positive momentum space wave packets and $\phi^r(\vec{k})$ denote the phase shifts of the corresponding modes.
Exploiting this decomposition, equation~\eqref{eqn:L1} becomes
\begin{align}
    \Delta L(t) &\approx - \frac{1}{2} \sum_{r \in \{+, \times\}} \int \mathrm{d}^3\vec{k} \; \psi_r(\vec{k}) \frac{A_{\parallel}^{r}(\varphi, \theta)}{\tilde{\omega}}
    \left\{\sin(\tilde{\omega} L_0)\cos(\omega t + \phi^r) + \left[\cos(\tilde{\omega} L_0) - 1\right]\sin(\omega t + \phi^r)\right\} \,, \label{eqn:L_final} 
\end{align}
where \(\theta\) and \(\varphi\) are the angles defining the relative orientation of the neutrino trajectory \(\gamma_0\) and the GW propagation direction. 
In terms of these angles we define the reduced GW frequency, \(\tilde{\omega}\), and the projected polarization tensors, \(A_{\parallel}^{r}(\varphi, \theta)\), as
\begin{align}
    &\tilde{\omega} := \omega (1-\cos(\theta)) \,,
    &&A_{\parallel}^{+}(\varphi, \theta) := \sin^2(\theta) \cos(2\varphi) \,,
    &&A_{\parallel}^{\times}(\varphi, \theta) := \sin^2(\theta) \sin(2\varphi) \,.
\end{align}
Depending on the experimental setup, the angles \(\theta\) and \(\varphi\) could change with time.
This can weaken the overall effect since it is only maximal for constant perpendicular neutrino and GW directions, i.e. \(\theta = \pi / 2\), see Appendix~\ref{app:EarthRotation} for a detailed discussion.

Next, we need to average over the duration of the data taking period \(T\), since neither production nor detection time are usually analyzed in neutrino experiments.
Thus, the oscillation probability is given by
\begin{align}
    P_{ab}(E, L_0) &= \frac{1}{T} \int_{0}^{T} \hat{P}_{ab}(E, L_0 + \Delta L(t)) \; \mathrm{d}t \equiv \langle \hat{P}_{ab}(E, L_0 + \Delta L) \rangle_{T} \,.
    \label{eqn:Time_average}
\end{align}
This gives rise to an additional damping of the oscillating terms in the oscillation probability \(\hat{P}\) which can be interpreted as a new source of decoherence.

Finally, we need to reconsider the physical meaning of our parameter 
\(L_0\). In the comparison of the prediction of the neutrino 
oscillation pattern with and without the GW effect, \(L_0\) plays 
the role of the physical distance of neutrino source and detector 
measured at a certain time \(t_0\). If a GW of any 
kind is present in the system this will replace the original value of 
the parameter \(L_0\) by \(\tilde{L}_0 = L_0 + \Delta L(t_0)\).
Therefore \(\tilde{L}_0\) is the actual, physical reference length 
entering the neutrino flavor transition probability implying that 
equation~\eqref{eqn:Time_average} needs to be evaluated at 
\(L_0 = \tilde{L}_{0} - \Delta L(t_0)\), corresponding to the 
subtraction of a constant off-set.
This subtlety is especially important if the GW does not oscillate over a full period during the time of data taking, \(T\), since then 
the parameter \(L_0\) is unphysical and the actual observable length 
is \(\tilde{L}_0\).
Hence, only time dependent changes in \(L(t)\) can be measured using 
the method described above.
From now on, we eliminate the unphysical parameter \(L_0\) in favor 
of the physical reference length \(\tilde{L}_0\) from our 
description and afterwards rename \(\tilde{L}_0\) to \(L_0\).

\begin{figure}
    \centering
    \includegraphics[width=\textwidth]{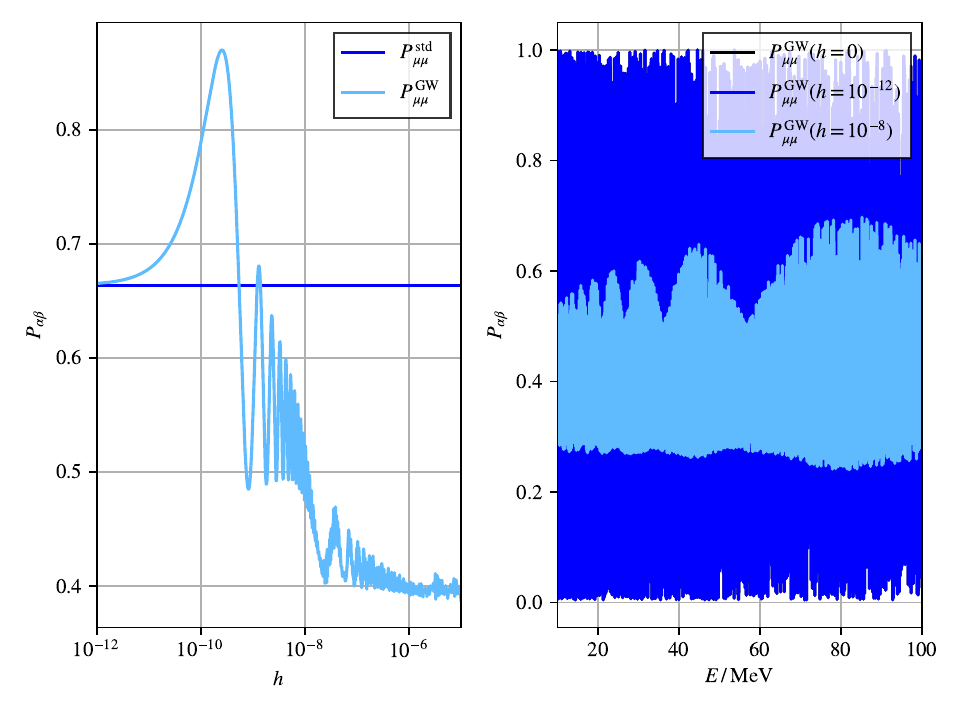}
    \caption{Muon neutrino survival probability, \(P_{\mu\mu}\), for different background GWs of different strains \(h\).
    Left: \(P_{\mu\mu}\) with (light blue) and without (blue) the influence of a GW for \(E = 10 \,\mathrm{MeV}\), \(L_0 = \SI{e10}{\kilo\metre}\).
    Right: \(P_{\mu\mu}\) for different fixed GW strains \(h \in \{0, 10^{-12}, 10^{-8}\}\) and variable energy at \(L_0 = \SI{e10}{\kilo\metre}\).
    The black oscillation curve (\(h = 0\)) almost coincides with the blue curve (\(h=10^{-12}\)) and is therefore hidden.}
    \label{fig:oscprob_GW_deco}
\end{figure}

In order to gain a better understanding of the impact of GWs on the neutrino oscillation probability,
we plot the \(\nu_\mu \to \nu_\mu\) survival probability in figure~\ref{fig:oscprob_GW_deco}.
For illustration purposes, we choose a baseline of \(L_0 = \SI{e10}{\kilo\metre}\), neutrino energies on the order of
\(\mathcal{O}(10\text{--}100\,\mathrm{MeV})\), standard neutrino mass and mixing parameters for normal mass ordering~\cite{Workman:2022ynf}
and an infinite wave packet width, i.e. we ignore decoherence from wave packet separation.
Furthermore, we consider a simple, plus polarized, plane GW with variable strain \(\psi_{+} \equiv h\) and frequency \(f \sim \SI{3e-7}{\hertz}\).
It is obvious that for small strains, the modified probability coincides with the standard expectation while for larger strains, we expect a significant deviation due to the averaging procedure just described.
The non-trivial, oscillating dependence on the strain can be explained by considering the mechanism of averaging the flavor transition probability over time.
As the strain increases, so does the amplitude of \(\Delta L\) and the averaging proceeds over many neutrino oscillation cycles resulting in the stable decoherence limit for large strains.
For strains from the intermediate region the time integral only produces averaging over a partial oscillation cycle leading to an average value varying with the strain.

\section{General Conditions for Observability of Gravitational Wave Induced Decoherence}
\label{sec:ObsCons}
In the following, we discuss the conditions that must be fulfilled for the new
decoherence effect to be observable in experiments:
\begin{enumerate}[(i)]
    \item Neutrino oscillations proceed sufficiently rapid such that for given GW parameters the deviation \(\Delta L\) is comparable to at least one of the oscillation lengths.
    \item Decoherence due to wave packet separation occurs later than GW induced decoherence, i.e. at larger baselines or smaller energies.
    \item The energy resolution or binning, \(\Delta E\), of the neutrino source and detector are sufficiently accurate such that oscillations are not averaged out, i.e. neutrino oscillations have to be sufficiently slow for the experiment to resolve them.
    \item The spatial extent of the region of neutrino origin must be much smaller than the oscillation length of the neutrinos to be resolved.
    \item Neutrino source and detector must not be fixed such that their physical distance can be influenced by the GW.
\end{enumerate}
Unfortunately, requirements one and three compete with each other:
On the one hand neutrino oscillations have to be fast enough such that the 
variation of the deviation \(\Delta L\) over the run time of the experiment 
is sufficiently large in order for the time average in 
equation~\eqref{eqn:Time_average} to damp the oscillations. On the other 
hand the oscillations must not be too fast since otherwise the finite energy bin width will 
damp the oscillations, and the effect due to the GW is unresolvable.
In the following we further quantify these loosely formulated criteria.
\paragraph{Criterion (i) and (ii)} First we
examine the maximal deviation $\Delta L$ from the baseline $L_0$.
Since we first want to explore the usefulness of neutrino oscillations as probes for coherent GW signals, 
we now restrict our discussion to plane GWs\footnote{That means GW signals with sufficiently peaked momentum distributions, i.e. \(\psi_r(\vec{k}) \approx a_r \delta^{(3)}(\vec{k} - \vec{k}_0)\)}.
In this case the distance correction $\Delta L$ from equation~\eqref{eqn:L_final} becomes
\begin{align}
    \Delta L(t) \approx -\frac{1}{2\tilde{\omega}} \sum_{r \in \{+, \times\}} a_{r} A_{\parallel}^{r}(\varphi, \theta)
    \left\{\sin(\tilde{\omega} L_0)\cos(\omega t + \phi^r) + \left[\cos(\tilde{\omega} L_0) - 1\right]\sin(\omega t + \phi^r)\right\} \,,
    \label{eqn:L_plane}
\end{align}
with mean frequency \(\omega\) and polarization amplitudes \(a_r\).
It is evident that for a fixed length \(L_0\) the amplitudes appearing in this expression oscillate with respect to the reduced frequency \(\tilde{\omega}\).
This makes it more difficult to set bounds on the high frequency region \(\tilde{\omega} \geq L_0^{-1}\) which is why, for the following discussion, we focus on the low frequency region \(\tilde{\omega} \ll L_0^{-1}\). 
In this region, we can express the maximal baseline deviation as
\begin{align}
    \max_{\omega t \in [0, 2\pi)} \vert \Delta L (t) \vert = \frac{h L_0}{2}\,,
\end{align}
with
\begin{align}
    h :\approx \sin^2(\theta) \max_{\omega t \in [0, 2\pi)} \vert a_+ \cos(2\varphi)\cos(\omega t + \phi^+) + a_\times \sin(2\varphi)\cos(\omega t + \phi^\times)\vert \,.
\end{align}
For a discernible effect, it is also crucial that the maximum deviation $\Delta L$
is comparable to the oscillation length $L^\text{osc}_{jk}$ of neutrinos:
\begin{equation}
    r = \frac{\max_{t\in\tau} \vert \Delta L (t) \vert}{L^\text{osc}_{jk}} = \frac{\Delta m_{jk}^2}{4 \pi E} \frac{h L_0}{2}\,,
\end{equation}
where we conservatively choose the ratio $r \approx 0.1$. The reason for this is that the $\nu_j$--$\nu_k$ interference
term needs to be sufficiently damped by the averaging procedure. 
For each scenario we consider the oscillation term with the largest mass difference $\Delta m_{jk}^2$,
since we are interested in the sensitivity of our framework.
However, the decision on which oscillation dominates must be 
made based on the experiment under consideration. 
From this we can derive a lower limit on the baseline \(L_\text{min}\) such that the effect becomes visible, i.e.
\begin{equation}
    L_\text{min} (E) = 8 \pi r \frac{E}{h \Delta m_{jk}^2} < L_0\,.
    \label{eqn:Condition_osc}
\end{equation}
On the other hand, it is essential for the baseline $L$ to be smaller than the 
coherence length $L_{jk}^\text{coh}$ of neutrinos to ensure that the effect of wave packet 
decoherence does not dominate and thus hide the desired effect. 
This results in an upper limit implying that the baseline $L_0$ has to vary within 
the following window to enable the observation of an effect: 
\begin{align}
    L_\text{min} (E) &< L_0 < L^\text{coh}_{jk} \\
    \Leftrightarrow 8 \pi r \frac{E}{h \Delta m_{jk}^2} &< L_0 < \frac{4 \sqrt{2} E^2}{\Delta m_{jk}^2} \sigma_x \,.
    \label{eqn:Condition_for_baseline}
\end{align}

\paragraph{Criterion (iii) and (iv)} The energy resolution $\Delta E$ must be sufficiently high to 
resolve the neutrino oscillations by the experiment. This implies that the frequency of 
the oscillation cannot be too large as too many wave cycles are contained within an energy 
bin, thus averaging out the effect. In the case of a sinusoidal oscillation of the form $\sin{\left(\omega t\right)}$,
this would mean that the period $\Delta t$, corresponding to the bin width, should be smaller than the inverse 
frequency: $\Delta t \lesssim \omega^{-1}$. This consideration can be applied to our case, leading to an 
energy resolution estimate for the resolution of the \(jk\)-oscillation:
\begin{align}
    \left\vert\Delta \left(\frac{L_0}{E}\right) \right\vert \lesssim \frac{2}{\Delta m_{jk}^2} 
    \quad \Leftrightarrow \quad 
    \Delta E \lesssim \frac{2}{\Delta m_{jk}^2}\frac{E^2}{L_0}\,.
    \label{eqn:Condition_for_energy_binning}
\end{align}
Furthermore, introducing the relative energy resolution $\delta E := \Delta E / E$,
we can express equation~\eqref{eqn:Condition_for_energy_binning} in the following way:
\begin{align}
    \label{eqn:Condition_for_energy_binning2}
    \delta E \lesssim \frac{L_{jk}^{\mathrm{osc}}}{2\pi L_0} =: \delta E_{\mathrm{thr}} \,. 
\end{align}
In addition to energy resolution, position resolution is also relevant. It is 
important that we know the origin of the neutrinos to a certain degree, 
hence the width of the production region $\Sigma_P$ must be smaller than 
the oscillation length $L^\text{osc}_{jk}$ to prevent additional averaging 
hiding the gravitationally induced effect:
\begin{equation}
    \Sigma_P \ll L^\text{osc}_{jk}\,.
    \label{eqn:Condition_for_source_uncertainty}
\end{equation}

\paragraph{Criterion (v)} In the measurement of GW effects, it is crucial to account for the 
absence of strong forces maintaining a constant physical distance between two reference points. 
For instance, in the case of LIGO, this is achieved through freely suspended mirrors~\cite{LIGOScientific:2016aoc}.
In our context, this demands the forces between the detector 
and the source to be negligible compared to the force of 
displacement induced by the GW.
Astrophysical neutrino sources do not pose a problem in this regard, as detector
and source are freely falling in the local gravitational fields.
One possibility would be to detect neutrinos from natural sources like the sun or, 
assuming they exist, decaying dark matter (DM) particles within the gravitational wells of celestial bodies.
At even larger distances pulsars, blazars or supernovae are possible candidates for neutrino sources.

\paragraph{Sensitivity Estimation for Coherent GW Signals at Neutrino Experiments}
With these criteria at hand one can now derive an estimate of the sensitivity of neutrino flavor oscillation experiments to the impact of GW on the oscillation pattern.
The most crucial observation in this regard is that criteria (i) and (iii) lead to a direct relationship between the minimal strain, 
\begin{align}
    h_{\mathrm{min}} := 2r \frac{L_{jk}^{\mathrm{osc}}}{L_0} \,,
\end{align}
such an experiment can be at most sensitive to, and its threshold relative energy resolution $\delta E_{\mathrm{thr}}$.
Combining this with equation~\eqref{eqn:Condition_for_energy_binning2}, we can infer that the minimal strain an experiment is sensitive to is bounded from below by its energy resolution
\begin{align}
    h_{\mathrm{min}} = 4 \pi r \delta E_{\mathrm{thr}} \,.
\end{align}
Assuming an optimistic energy resolution $\delta E \approx \delta E_{thr}$ on the percent level~\cite{JUNO:2020xtj} together with our 
choice of $r \approx 0.1$ for a realistic neutrino experiment this leads to
\begin{align}
    h \gtrsim 4 \pi r \delta E_{\mathrm{thr}} \sim 0.013 \,,
    \label{eqn:hmin}
\end{align}
which is several orders of magnitude larger than sensitivities of current laser interferometer experiments, e.g. LIGO, probing strains on the order of $h \sim 10^{-20}$ and below.
To achieve this level of sensitivity using neutrino oscillation based experiments a relative energy resolution of at least $\delta E \sim 10^{-18}\%$ is needed.

Even if at future neutrino telescopes a tremendous improvement in relative energy resolution would be achieved,
the effect of wave packet decoherence and also averaging over a potentially macroscopic neutrino production region
imposes further restrictions on the sensitivity of neutrino oscillations to coherent GW signals.
These effects are especially relevant for neutrinos of astrophysical origin representing the most 
realistic candidates for the observation of GW induced decoherence according to criteria (i) and (v).

An additional difficulty arises from the duration of coherent GW signals, $T_\text{GW}$,
compared to the data taking period of the neutrino experiment $T_\text{exp}$.
If $T_\text{GW} \ll T_\text{exp}$ the averaging effect will be washed out and the sensitivity decreases.
Based on this consideration one could only expect neutrino based GW detection to be sensitive to astrophysical
events emitting GW signals of durations of $\mathcal{O}\left(\num{10} \,\text{yr}\right)$
or greater, like supermassive blackhole binary (SMBHB)~\cite{Sesana:2013dma} mergers or
extreme-mass-ratio inspirals (EMRI)~\cite{Ye:2023uvh}.
It is worth noting that SMBHB mergers typically last longer than 10 years while 
emitting GWs with a constant frequency~\cite{Chen:2018znx}.
Taking all criteria into account, observing any significant effect from coherent GWs on neutrino oscillations
is rather unrealistic. The robustness of the developed criteria is further validated in appendix~\ref{app:num} with a 
statistical toy analysis taking all the discussed effects numerically into account.

Therefore, in the next section, we consider another scenario namely the detection of the SGWB using neutrino flavor oscillations.
Even after the preceding discussion this still remains an interesting opportunity since the mechanism by which decoherence 
is induced in the neutrino system is very different from the mechanism discussed so far.

\section{Parameter Constraints for the Stochastic Gravitational Wave Background}
\label{sec:sgw}
Until now, we considered the interplay of neutrino oscillations and plane GW signals of well defined origin.
Another interesting phenomenon can be studied in the $\si{\nano\hertz}$ frequency band, i.e. the SGWB.
In parallel to the recent discovery of the SGWB by PTA collaborations, already several works about possible consequences
for neutrino oscillations have been published~\cite{Dvornikov:2021sac, Lambiase:2023pxd}.
While in these studies the interplay of GW induced decoherence and other sources of oscillation damping 
was already discussed qualitatively, 
we now adopt a more quantitative approach and also analyze the scenarios
considered in reference~\cite{Lambiase:2023pxd} statistically.

To this end, we revisit equation~\eqref{eqn:prob0} and incorporate not only the wave packet separation effect into the 
decoherence term, but also the damping term induced by the SGWB,
\begin{equation}
    \mathcal{D}_{jk} (E,L_0) = \left(\frac{L_0}{L^\text{coh}_{jk}}\right)^2 + \Gamma_{jk}^\text{SGWB} (E)\,.
\end{equation}
This approach, while being different to the one employed so far, roots in very similar considerations of the modification
of the neutrino propagation due to the GW in the density matrix formalism, cf. reference~\cite{Dvornikov:2021sac} for further details.
For the following analysis, we do not average over the running time of the neutrino experiment anymore as we have shown that this effect is negligible.
Therefore, we only take into account the stochastic perturbations of the neutrino trajectories caused by an ensemble of GWs through which the neutrino system propagates.

The expression for $\Gamma_{jk}^\text{SGWB}$ according to reference~\cite{Lambiase:2023pxd}, is given by
\begin{equation}
    \Gamma_{jk}^\text{SGWB}(E, L_0) =\left(\frac{3 H_0}{8 \pi L^\text{osc}_{jk}}\right)^2 
    \int_{f_{\mathrm{min}}}^{f_{\mathrm{max}}} \frac{\mathrm{d}f}{f^5} \sin^2(\pi f L_0) \Omega_{\mathrm{GW}}(f)\,,
\end{equation}
where $\Omega_{\mathrm{GW}}$ is the fractional energy density of the SGWB, $H_0$ is the Hubble rate today, and $f_{\mathrm{min}}$ and $f_{\mathrm{max}}$ are the minimum and maximum frequencies of the SGWB spectrum, respectively.
As described in reference~\cite{Lambiase:2023pxd}, the SGWB frequency distribution, $h_c$, is modelled by a power-law spectrum of the form
\begin{equation}
    h_c (f) = A_* \left(\frac{f}{f_\text{yr}}\right)^{\frac{3-\gamma}{2}}\,.
\end{equation}
Here $f_\text{yr} = 1 \text{yr}^{-1} \approx \SI{31.7}{\nano\hertz}$ is the reference
frequency, $\gamma$ is the spectral index, $f_\text{min}$ is the lowest frequency of the GW spectrum
and $A_*$ is the amplitude.
The resulting fractional energy density reads
\begin{align}
    \Omega_{\mathrm{GW}}(f) &= \frac{2 \pi^2}{3 H_0^2} f_\text{yr}^2 \vert A_{\ast} \vert^2 \left(\frac{f}{f_\text{yr}}\right)^{1-\gamma} \,.
\end{align}
Given this frequency distribution the UV contribution of the GW spectrum is negligible and we can take the limit $f_{\mathrm{max}} \to \infty$, as has been done in reference~\cite{Lambiase:2023pxd}.
In this limit, the SGWB decoherence term becomes
\begin{align}
    \Gamma_{jk}^\text{SGWB} = \frac{3}{64} \left(\frac{\vert A_* \vert}{f_\text{yr} L^\text{osc}_{jk}}\right)^2 \left(\frac{f_\text{min}}{f_\text{yr}}\right)^{1-\gamma} \left(\frac{1}{\left(\gamma - 1\right)} - \mathrm{Re}\left(E_{\gamma}(2\pi i f_{\mathrm{min}}L_0)\right)\right) \,,
\end{align}
where $E_{\gamma}(z)$ is the exponential integral\footnote{The exponential integral is defined as $E_{\gamma}(z) := \int_{1}^{\infty} e^{-t} t^{-\gamma} \;\mathrm{d}t$.} of order $\gamma$.
Therefore, in total, this model has three free parameters:
The minimum frequency $f_{\mathrm{min}}$, the amplitude of the spectrum $A_{\ast}$ and the spectral index $\gamma$.

%
First, we determine in section~\ref{ssec:sgwb_theory} the neutrino energies and their origins fulfilling the criteria 
developed in section~\ref{sec:ObsCons} in order to see the effect of the SGWB on the neutrino oscillations. 
Afterwards, we explain the employed statistical analysis and the toy eperiment 
setup in section~\ref{ssec:sgwb_stat_ana} used to arrive at the results presented in section~\ref{ssec:SM_pulsar_SGWB}.

\subsection{Neutrinos from Galactic Pulsars}
\label{ssec:sgwb_theory}
In this section we will discuss possible neutrino sources ranging from neutrinos from long baseline neutrino 
experiments to neutrinos from astrophysical origin in order to decide which offer the best opportunities to 
observe the effects of the SGWB in neutrino flavor oscillations.
To this end, we can again go back to criteria (ii) to (v) from section~\ref{sec:ObsCons} since these are generally applicable
for the observation of decoherence from GW effects in flavor oscillations.

Beginning with long baseline neutrino experiments, we find that criterion (v) is not fulfilled and it is therefore 
conceptionally not possible to detect the SGWB effect in these experiments. The same holds for atmospheric neutrinos.
Going to higher baselines the next neutrino source is the sun, which produces neutrinos in the $100 \si{\kilo\electronvolt}$ to $\si{\mega\electronvolt}$
energy range. One challenge lies in the requirement that the neutrino production region needs to be sufficiently small 
compared to the oscillation length $L^\text{osc}_{jk}$, cf. equation~\eqref{eqn:Condition_for_source_uncertainty},
in order to detect the decoherence effect caused by GWs.
The ratio $L^\text{osc}_{jk} / \Sigma_P$ scales linearly with the energy $E$.
Consequently, the higher the energy of the neutrinos, the larger the neutrino production region can be 
without significantly averaging the oscillations.

Table~\ref{tab:uncertainty_of_source_location} gives the oscillation lengths, $L_{jk}^{\mathrm{osc}}$, of neutrinos of various energies, $E$, and thereby provides an intuition about how accurate the region of origin would need to be known.
\begin{table}
    \caption{Values for the oscillation length $L^\text{osc}_{jk}$ for different neutrino energies $E$ and mass splitting of $\Delta m_{jk}^2 \sim \SI{e-3}{\electronvolt\squared}$.}
    \begin{center}
    \begin{tabular}{c c} 
    \toprule
    $E$ & $L^\text{osc}_{jk} \, / \, \si{\kilo\meter}$\\
    \midrule
    $\si{\kilo\electronvolt}$ & $\num{2.4e-3}$\\
    $\si{\mega\electronvolt}$ & $\num{2.4}$\\
    $\si{\giga\electronvolt}$ & $\num{2.4e3}$\\
    $\si{\tera\electronvolt}$ & $\num{2.4e6}$\\
    $\si{\peta\electronvolt}$ & $\num{2.4e9}$\\
    $\si{\exa\electronvolt}$ & $\num{2.4e12}$\\
    \bottomrule
    \end{tabular}
    \label{tab:uncertainty_of_source_location}
    \end{center}
\end{table}
We see that for $\si{\kilo\electronvolt}$ to $\si{\mega\electronvolt}$ neutrinos from the sun the origin has to be known up to $\mathcal{O}(\si{\meter})$ to $\mathcal{O}(\si{\kilo\meter})$. 
As the neutrinos get produced in the core of the sun this condition is of course not met. Another possible neutrino source are, 
assuming they exist, decaying dark matter particles within the gravitational wells of celestial bodies. Since no 
neutrinos possibly coming from such sources have been identified, this is not a promising scenario.

After pointing out that the prospects to probe the SGWB effect on neutrino oscillations at distances within our solar system are rather bleak, we now consider astrophysical 
neutrino sources outside the solar system.

A possible astrophysical source for neutrinos are supernovae~\cite{Janka:2017vlw} typically
emitting $\si{\mega\electronvolt}$ neutrinos~\cite{Suliga:2022ica}. However, since we would need to
know the location of the neutrino source within $\si{\kilo\meter}$ range to identify decoherence from GWs no 
sensitivity is expected for such sources, see table~\ref{tab:uncertainty_of_source_location}.

Another astrophysical source of neutrinos are blazars~\cite{Oikonomou:2022gtz}, i.e. active galactic nuclei (AGNs) with
jets pointing in our direction. Neutrinos from such a source are expected to be in the energy range 
from $\si{\peta\electronvolt}$ to $\si{\exa\electronvolt}$~\cite{Boettcher:2022dci}. 
According to table~\ref{tab:uncertainty_of_source_location}, for such high-energy neutrinos 
the spatial size of the neutrino source can be as large as approximately $\SI{e14}{\kilo\meter}$.
While this may be possible for neutrinos produced in the vicinity of a blazar, 
these sources still do not appear to be suitable candidates to measure the effects studied in this work.
This is due to the expectation of a low flux of high energy neutrinos reaching the Earth resulting in a non-significant 
neutrino count, cf. reference~\cite{Oikonomou:2019djc}, compared to the required event count of 
$N = \num{e4} - \num{e5}$ in our set-up.

Thus, the most promising astrophysical sources of neutrinos are the jets of pulsars~\cite{Protheroe:1997er,Gandhi:2000kq,Beall:2001pa,Guetta:2002hv,Guetta:2002du,Razzaque:2002kb,Dermer:2003zv,Bednarek:2003cv,Link:2004yn,Link:2006pd,Fang:2014qva} from our own Galaxy,
like for example the Vela (\(L_0 = 294^{+76}_{-50}\,\mathrm{pc}\)~\cite{Caraveo:2001ud}), Crab (\(L_0 = 1.9^{+0.22}_{-0.18}\,\mathrm{kpc}\)~\cite{Lin:2023erb}) 
or Cas A (\(L_0 = 3.5 \pm 0.2\,\mathrm{kpc}\)~\cite{Pavlov:2003eg}) pulsars.
Due to the relatively high expected neutrino fluxes of about \(\mathcal{O}(30 \, \mathrm{km}^{-2}\mathrm{yr}^{-1})\), 
event counts of \(N \sim \num{e4}\) seem to be realistic at future \(100\,\mathrm{km}^2\)--neutrino telescopes 
and a data taking period of about \(T_{\mathrm{exp}} \sim 20\,\mathrm{yr}\).
Estimating the size of the region where neutrinos originate from a pulsar involves 
considering various theoretical models and observational constraints. 
According to these scenarios, very high energy neutrinos may be generated in the vicinity of the pulsar's magnetosphere 
and in the surrounding pulsar wind nebula, where protons are accelerated to high energies and interact with photons or 
other protons~\cite{Protheroe:1997er,Gandhi:2000kq,Beall:2001pa,Guetta:2002hv,Guetta:2002du,Razzaque:2002kb,Dermer:2003zv,Bednarek:2003cv,Link:2004yn,Link:2006pd,Fang:2014qva,Guepin:2019fjb,Asthana:2023vvk} producing pions which in turn decay into neutrinos.
The magnetosphere is estimated to be of the order of several kilometers, depending on the specific properties of the pulsar~\cite{Guepin:2019fjb}.
Furthermore, we expect neutrino energies in the energy range $E \in \left[\SI{100}{\tera\electronvolt}, \SI{2}{\peta\electronvolt}\right]$,
as predicted e.g. in reference~\cite{Link:2006pd}.
Thus, the width of the production region is likely to be sufficiently small compared to the oscillation length, 
cf. table~\ref{tab:uncertainty_of_source_location}. 
Even in case the oscillation length is not much smaller than the size of the production region but $L^\text{osc}_{jk} \lesssim \Sigma_P$ applies, galactic pulsars may still remain suitable sources with a potential 
additional oscillation damping originating from a necessary averaging over the neutrino production region.

\subsection{Statistical Analysis and Toy Experiment Setup}
\label{ssec:sgwb_stat_ana}
In order to estimate the actual sensitivity of neutrino oscillations on the presence of the SGWB,
we perform a toy analysis on specific scenarios motivated by the considerations presented in reference~\cite{Lambiase:2023pxd}
taking into account all deviations from the standard oscillation pattern.
To this end, we model a hypothetical experimental set-up and estimate the spectral amplitude and spectral index regions it would be sensitive to.
Each of these hypothetical experiments is assumed to count neutrino event rates in finite, equal width energy bins and 
the corresponding neutrino signal is expected to originate from a source at some average distance \(L_0\).
In all scenarios we assume a power law energy spectrum for neutrinos from 
astrophysical environments where charged particles are accelerated 
in jets and interact with each other or the environment to produce neutrinos.

The number of neutrino events is then predicted according to the probability density,
\begin{align}
    \rho_{b}(E) &= \sum_{a} \varphi_{a}(E) P_{a b}(E, L_0) \,,
\end{align}
where \(\varphi_a\) is the neutrino flavor and energy distribution at the source and \(P_{ab}\) is
the flavor transition probability of the given model.
We assume a neutrino flux equivalent to a power-law spectrum \(\varphi \propto E^{-2}\).

The likelihood of our number counting experiment is given by
\begin{equation}
    \mathcal{L}(A_{\ast}, \gamma, f_{\mathrm{min}} \; \vert \; \vec{n}) 
    = \prod_{b = 1}^{n_f} \prod_{i=1}^n \text{Pois} (n_{ib}, \eta_{ib} (A_{\ast}, \gamma, f_{\mathrm{min}}))\,,
    \label{eqn:NLLR}
\end{equation}
where $\eta_{ib}(A_{\ast}, \gamma, f_{\mathrm{min}})$ denotes the theoretically expected value of events in bin $i$ and flavor $b$
depending on the spectral amplitude $A_{\ast}$, spectral index $\gamma$ and the low frequency cut-off $f_{\mathrm{min}}$,
and $n_{ib}$ denotes the event count measured at an experiment.
We generate a toy data set, assuming the expected number of events for each 
energy and flavor bin $n_{ib}$ to follow the flat spacetime prediction with $\Gamma_{jk}^{\mathrm{SGWB}}(E, L_0) \equiv 0$.
In order to reproduce the analysis from reference~\cite{Lambiase:2023pxd},
we choose several fixed values for the low frequency cut-off $f_{\mathrm{min}}$ and
perform a likelihood ratio test on the remaining two dimensional parameter set in the $A_{\ast}$--$\gamma$ plane.
To this end, we evaluate the negative logarithmic likelihood ratio (NLLR),
\begin{equation}
    \Lambda(\vec{n}) = -2 \ln{\left(\frac{\mathcal{L}({A_{\ast}}_0, \gamma_0, f_{\mathrm{min}}^{\mathrm{fix}})}{\sup_{(A_{\ast}, \gamma) \in \mathbb{R}^2} \left(\mathcal{L}(A_{\ast}, \gamma, f_{\mathrm{min}}^{\mathrm{fix}})\right)}\right)} \,,
\end{equation}
on a grid in a region of our parameter space, where $({A_{\ast}}_0, \gamma_0)$ denotes the corresponding grid point.
Further details of the setup and statistical analysis can be found in appendix~\ref{app:Statistics}.

\subsection{Probing the Stochastic Gravitational Wave Background with Neutrinos from Galactic Pulsars}
\label{ssec:SM_pulsar_SGWB}
Despite the discussion in section~\ref{ssec:sgwb_theory} regarding the possible neutrino sources, 
we refrain from using the exact parameter configuration favored there. 
Pulsars in a realistic distance to 
Earth emitting neutrinos in the low $\si{\peta\electronvolt}$ energy range can not fulfill criterion (iii) if 
we demand an energy resolution at the sub-percent level. In order to resolve the oscillations from such sources, e.g. the Vela Pulsar, 
neutrino oscillation experiments would need an energy resolution of $\delta E \sim 10^{-8}$. 
Therefore, we focus on the scenario from reference~\cite{Lambiase:2023pxd}, where neutrinos are emmited from a pulsar in the energy range 
\(E \in [1\,\mathrm{PeV}, 100\,\mathrm{PeV}]\) in a distance of $L_0 \approx 50 \,\text{lyr}$ as can be 
derived from the choice of the relative energy resolution of $\delta E \sim 0.1\%$ at $E \sim 100\,\mathrm{PeV}$.
However, we do not expand $\Gamma^\text{SGWB}_{jk}$ for 
$L_0 \gg 1 / f_\text{min}$ as has been done in reference~\cite{Lambiase:2023pxd} for different scenarios 
since this expansion does not hold for the present parameter configuration. 
In order to be able to observe a significant deviation from the standard 
case, we assume a total number of $N \approx \num{2e4}$ events and a flat $\varphi \propto E^{-2}$ flux spectrum. 
Furthermore, we assume a pure initial $\nu_\mathrm{e}$ flux as in reference~\cite{Lambiase:2023pxd} and to 
incorporate effects from wave packet decoherence, we also consider wave packet 
widths of $\sigma_x \sim \SI{1}{\nano\metre}$.
\begin{figure}
    \centering
    \includegraphics{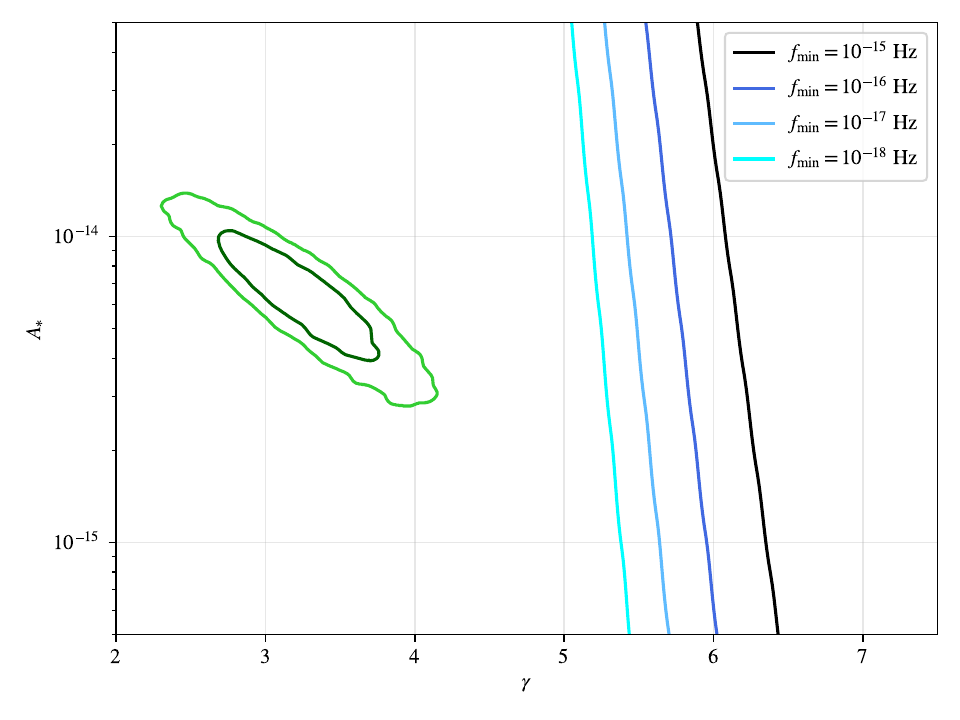}
    \caption{Shown in dark (light) green are the $68\,\%$ ($95\,\%$) credible region of the NANOgrav result for the 15-year data taking period~\cite{NANOGrav:2023gor}. 
    The other lines correspond to the limits at 68\,\% CL on the GW amplitude \(A_{\ast}\) and the spectral index \(\gamma\) for different values 
    of minimum frequencies, $f_\text{min} \in \{10^{-18},10^{-17},10^{-16},10^{-15}\}$.
    Parameter configurations to the upper right of these lines are excluded if the effect of the SGWB is not observed
    in neutrino oscillation experiments at baselines \(L_0 \sim 50\,\mathrm{lyr}\), energies \(E \in [1\,\mathrm{PeV}, 100\,\mathrm{PeV}]\), bin widths \(\Delta E \lesssim 0.1 \mathrm{PeV}\) and neutrino wave packet widths \(\sigma_x \sim 1\,\mathrm{nm}\).}
    \label{fig:SGWB_PeV}
\end{figure}

The results of our analysis are visualized in figure~\ref{fig:SGWB_PeV}.
In dark (light) green, we show the $68\,\%$ ($95\,\%$) credible region of the NANOgrav result for the 15-year data taking period~\cite{NANOGrav:2023gor}.
The other lines correspond to the \(68\,\%\) confidence level contours of the NLLR function in the \(\gamma\)--\(A_{\ast}\) plane for different values of minimum frequencies, $f_\text{min}$, respectively.
The parameter regions to the upper right of these contours are excluded for a data set generated assuming the absence of a SGWB.
For the corresponding configurations, i.e. higher strain and bigger spectral index than indicated by the contour, neutrino oscillations are damped compared to the prediction in flat spacetime.
The qualitatively behavior of our results are in agreement with those from 
reference~\cite{Lambiase:2023pxd}.
In reference~\cite{Lambiase:2023pxd} the exclusion limits have been determined by setting $\Gamma^{\text{SGWB}}_{jk} \gtrsim 1$.
The bounds we obtain are significantly weaker compared to those derived in reference~\cite{Lambiase:2023pxd}, where they expand $\Gamma^\text{SGWB}_{jk}$ 
for $L_0 \gg 1 / f_\text{min}$, since the energy resolution and distance are not further specified in the first part of their analysis. However, 
the parameter configuration considered in this work accounts for energy resolution on the sub-percentage level and due to the resulting distances one has to
take into account the full expression.
This yields a significant cancellation between the leading term of the expansion and the remaining expression resulting in the loss
of any sensitivity of neutrino oscillation experiments to the SGWB, cf. figure~\ref{fig:SGWB_PeV}.
It is important to note, this analysis relies on the assumption, that there is a suitable source
in this distance emitting a large number of neutrinos in the desired energy range.
These dismal prospects are further substantiated by the fact that once decoherence due to the SGWB occurs, the difference
in this toy scenario to the ``standard'', i.e. no SGWB, case corresponds to $\approx1.5\,\sigma$ which is below any threshold
to be regarded as significant. Moreover the allowed regions we obtain are already excluded by CMB data, see e.g. reference~\cite{Namikawa:2019tax}. 

For much larger cut-off frequencies like \(f_{\mathrm{min}} \in [10^{-12}, 10^{-9}]\,\si{\hertz}\) on the other hand, neutrinos with energies in the \(\mathrm{MeV}\) range would be needed in order to observe a significant effect from the SGWB.
A possible source for astrophysical neutrinos of this energy would be supernova neutrinos where typical wave packet widths are \(\sigma_x \sim \SI{0.1}{\pico\meter}\)~\cite{Kersten:2015kio}.
In this case, as discussed in reference~\cite{Lambiase:2023pxd}, decoherence from wave packet separation has already occured 
and no effect from the SGWB can be expected unless coherence is restored, e.g. by matter effects in the so called catch-up mechanism~\cite{Kersten:2015kio}.
Even if coherence would be restored by any effect, it is questionable, if the neutrino oscillations could be resolved, since that would require an energy resolution or binning of
\begin{align}
    \Delta E \sim 4.17 \times 10^{-12} \left(\frac{10^{-5}\,\mathrm{eV}^2}{\Delta m_{jk}^2}\right)
    \left(\frac{E}{\mathrm{MeV}}\right)^2
    \left(\frac{\mathrm{lyr}}{L}\right) \, \mathrm{MeV} \,,
\end{align}
cf. equation~\eqref{eqn:Condition_for_energy_binning}.
In addition, uncertainties in the baseline of astrophysical neutrinos can also affect the constraints on GW parameters.

\section{Conclusions}
\label{sec:conc}
In this paper we have studied the influence of GW signals on neutrino 
oscillations.
We have developed a heuristic yet robust model describing the propagation 
of neutrinos in a spacetime curved by GWs.
After pointing out how GWs influence neutrino flavor oscillations by altering the propagation length of neutrinos,
we have systematically established conditions under which this effect is observable. 
As it turns out the minimal strain of a coherent GW detectable at neutrino experiments is bounded from 
below by the energy resolution of the experiment. As today's and near future experiments have an energy 
resolution at best at the percent order, this minimal strain is on the order of $\mathcal{O}(10^{-2})$ which is many 
orders of magnitude above any known source of GWs. To compete with laser interferometer experiments an 
improvement of the energy resolution by 18 orders of magnitude would 
be required. Therefore 
it is very unrealistic to observe coherent GW signals with neutrino oscillations.

Futhermore, we have examined the influence of the SGWB on neutrino oscillations. To this end we have argued 
that neutrinos orignating from pulsars are the most promising candidates to measure the effect of the SGWB. Nevertheless, 
for realistic pulsars the energies of the neutrinos combined with the corresponding baseline results in an energy resolution of $\mathcal{O}(10^{-8})$,
corresponding to an improvement of 7 orders of magnitude compared to current and planned experiments. Therefore we studied a toy source 
in accordance with reference~\cite{Lambiase:2023pxd} who did a first phenomenological analysis on this topic.
We improved on previous investigations of this scenario that 
have only qualitatively discussed the effect of wave packet decoherence.
When considering wave packet decoherence and the finite energy resolution 
and binning of realistic experiments,
we show explicitly that $\si{\mega\electronvolt}$ neutrinos are not sensitive to the SGWB,
while $\si{\peta\electronvolt}$ neutrinos from the dicussed sources might provide in very peculiar scenarios
a possibility to study the SGWB.
It should, however, be emphasized that our toy example corresponds to a very idealized scenario 
assuming $\mathcal{O}(10^4)$ event counts at very high energies of up to $\SI{100}{\peta\electronvolt}$ at distances 
of $\mathcal{O}(50\,\text{lyr})$ and with a sub-percent level energy resolution. This cannot be expected to be
achieved in the near future.

The scenario of 
heavier neutrino mass eigenstates is not discussed since the lower limit accessible on the strain in equation~\eqref{eqn:hmin} is
independent of the mass splitting. For the same reason meson mixing is not a promising alternative to detect GWs.
 
Another possible approach to measure the influence of GWs on neutrino oscillations would be to design an experiment 
detecting neutrinos precisely at energies on the verge of wave packet decoherence.
Instead of introducing more decoherence into the system the influence of a GW could indeed also restore the coherence by shifting the baseline, accordingly.
A possible experimental set-up could involve two phases of data taking where for example in the first phase decoherence is observed, while oscillations are restored in the second phase.
In this case one could conclude that a GW of frequency \(f \sim T_{\mathrm{exp}}\) has passed through the neutrino trajectory.
However, since this approach depends strongly on the phase of the GW only limited information can be expected.

In summary, we have discussed how neutrino oscillations are affected by GW signals.
Despite optimistic claims in the literature, neutrino oscillations are currently not suited to probe GW signals, as the energy 
resolution of neutrino oscillation experiments are too low to resolve any kind of effect.
They suffer from relatively low sensitivities especially compared to laser interferometry experiments,
which is mostly due to the interplay of the GW induced decoherence effect and other sources of oscillation averaging,
especially the energy resolution of neutrino oscillation experiments.

\newpage
\appendix

\section{Geodesics in Gravitational Wave Spacetimes: The Boundary Value Problem}
\label{app:Geodes}

Within our heuristic approach, we replace the baseline with a modified length, as it is subject to the effects of GW as can be seen in equation~\eqref{eqn:Length_metric}.
In order to calculate this spatial length, we first rewrite equation~\eqref{eqn:Length_metric} by using the on-shell condition
\begin{align}
    1 &= g_{\mu\nu} \dot{\gamma}^{\mu} \dot{\gamma}^{\nu} \\
    \Leftrightarrow 1 &= (\dot{\gamma}^{0})^2 + g_{jk} \dot{\gamma}^{j} \dot{\gamma}^{k} \\
    \Leftrightarrow -g_{jk} \dot{\gamma}^{j} \dot{\gamma}^{k}&= (\dot{\gamma}^{0})^2 - 1 \,.
\end{align}
Here we furthermore employed the transverse traceless (TT) gauge implying \(g_{j0} \equiv 0\) and \(g_{00} \equiv 1\) in our convention.
Substituting this result into equation~\eqref{eqn:Length_metric} yields
\begin{equation}
    L = \int_{\tau_P}^{\tau_D} \left(\left(\dot{\gamma}^0\right)^2 - 1\right)^{\frac{1}{2}}\text{d}\tau\,.
    \label{eqn:Length_with_Geodesic}
\end{equation}
To calculate this modification, it is initially necessary to compute the neutrino geodesic in our GW spacetime. 
To this end, we solve the geodesic equation for given boundary conditions (BC), i.e.
\begin{align}
    \ddot{\gamma}^\mu (\tau) &= -\Gamma^\mu_{\alpha\beta} (\gamma(\tau)) \dot{\gamma}^\alpha(\tau) \dot{\gamma}^\beta(\tau) \,,\\
    \gamma^{\alpha}(\tau_P) &= X_P^{\alpha} \,,\label{eqn:BC1}\\
    \gamma^{\alpha}(\tau_D) &= X_D^{\alpha} \,.\label{eqn:BC2}
\end{align}
Since our metric is close to \(\eta = \mathrm{diag(1,-1,-1,-1)}\), we adopt the ansatz
\begin{align}
    \gamma^{\alpha}(\tau) = \gamma_0^{\alpha}(\tau) + \epsilon \Delta \gamma^{\alpha}(\tau) \,.
\end{align}
In general the boundary value problem (BVP) of the geodesic equation is not well posed for arbitrarily far separated production and detection points since there may be more than one shortest curve connecting these points.
In our specific scenario, however, we assume that geodesics between two points are inherently 
unique because our spacetime closely resembles Minkowski spacetime.

Upon substituting this ansatz into the geodesic equation, we obtain, keeping only terms up to first order in \(\epsilon\),
\begin{equation}
    \ddot{\gamma}_0^{\mu} (\tau) + \epsilon \Delta \ddot{\gamma}^\mu (\tau)
    \approx -\frac{\epsilon}{2} \eta^{\mu\nu} \left(\partial_\alpha h_{\nu\beta} + \partial_\beta h_{\nu\alpha} - \partial_\nu h_{\alpha\beta}\right)\vert_{\gamma_0(\tau)} \dot{\gamma}_{0}^{\alpha} \dot{\gamma}_{0}^{\beta} \,.
\end{equation}
By matching terms of the same order in \(\epsilon\) we arrive at the following set of equations:
\begin{align}
    \ddot{\gamma}_0^{\mu} (\tau) &= 0 \label{eqn:Zeroth_Order_GE} \,, \\
    \Delta \ddot{\gamma}^\mu(\tau) &= -\frac{\eta^{\mu\nu}}{2} 
    \left(\partial_\alpha h_{\nu\beta} + \partial_\beta h_{\nu\alpha} - \partial_\nu h_{\alpha\beta}\right) \vert_{\gamma_0(\tau)} \dot{\gamma}_{0}^{\alpha} \dot{\gamma}_{0}^{\beta} \,.
\end{align}
Equation~\eqref{eqn:Zeroth_Order_GE} is easily solved by \(\gamma_0^{\mu}(\lambda) = a^\mu + \lambda u^\mu\).
Using our BC from above (at zeroth order in \(\epsilon\)) on the flat spacetime part of the geodesic, the constants of integration can be determined to yield 
\begin{align}
    \gamma_0^\mu (\tau) = X_P^\mu + \left(\tau - \tau_P\right) u^\mu\,,
\end{align}
where $u^\mu = (X_D^\mu - X_P^\mu) / (\tau_D - \tau_P)$.
For the geodesic correction \(\Delta \gamma\), we hence obtain the differential equation
\begin{equation}
    \Delta \ddot{\gamma}^{\mu}(\tau) = - \frac{\eta^{\mu\nu}}{2} \left(2\partial_\alpha h_{\nu\beta} - \partial_\nu h_{\alpha\beta}\right)\vert_{\gamma_0(\tau)} u^\alpha u^\beta =: \Delta a^\mu (\tau)\,,
\end{equation}
and BC, \(\Delta \gamma(\tau_P) = \Delta \gamma(\tau_D) = 0\), at first order in \(\epsilon\).
Integrating \(\Delta \ddot{\gamma}^{\mu}(\tau)\) leads to the correction to the four velocity, i.e.
\begin{equation}
    \Delta \dot{\gamma}^\mu (\tau) = \Delta u_P^\mu + \int_{\tau_P}^\tau \Delta a^\mu (\tau^\prime) d\tau^\prime\,.
\end{equation}
Integrating once again finally results in 
\begin{align}
    \Delta \gamma^\mu (\tau) = \Delta \gamma_P^\mu + \Delta u_P^\mu (\tau_D - \tau_P) 
    + \int_{\tau_P}^{\tau} \int_{\tau_P}^{\tau^{\prime}} \Delta a^\mu (\tau^{\prime\prime}) \;\mathrm{d}\tau^{\prime\prime} \mathrm{d}\tau^\prime \,.
\end{align}
By using our BC from equation~\eqref{eqn:BC1} and~\eqref{eqn:BC2}, we are able to determine both constants of integration, i.e. \(\Delta \gamma_P^\mu\) as well as \(\Delta u_P^\mu\):
\begin{align}
    0 &\stackrel{!}{=} \Delta \gamma^\mu (\tau_P) = \Delta \gamma_P^\mu \,,\\
    0 &\stackrel{!}{=} \Delta \gamma^\mu (\tau_D) = \Delta u_P^\mu \left(\tau_D - \tau_P\right)\ + \int_{\tau_P}^{\tau_D}\text{d} \tau^\prime \int_{\tau_P}^{\tau^\prime} \text{d} \tau^{\prime \prime} \Delta a^\mu (\tau^{\prime\prime})\,, \\
    \Leftrightarrow \Delta u_P^\mu &= - \Delta \bar{u}^{\mu} = \frac{-1}{\tau_D - \tau_P} \int_{\tau_P}^{\tau_D} \text{d} \tau^\prime \int_{\tau_P}^{\tau^\prime} \text{d} \tau^{\prime\prime} \Delta a^\mu (\tau^{\prime \prime}) \,.
\end{align}
Here we also defined $\Delta \bar{u}^\mu$, i.e. the average velocity correction between $X_P$ and $X_D$.
In total, the solution is given by
\begin{align}
    \gamma^{\mu}(\tau) &= u^{\mu} (\tau - \tau_P) + X_P^{\mu} 
    - \frac{\tau - \tau_P}{\tau_D - \tau_P} \int_{\tau_P}^{\tau_D}\int_{\tau_P}^{\tau^{\prime}} \Delta a^{\mu}(\tau^{\prime\prime}) \;\mathrm{d}\tau^{\prime\prime}\mathrm{d}\tau^{\prime}
    + \int_{\tau_P}^{\tau}\int_{\tau_P}^{\tau^{\prime}} \Delta a^{\mu}(\tau^{\prime\prime}) \;\mathrm{d}\tau^{\prime\prime}\mathrm{d}\tau^{\prime} \,, \\
    \Delta a^{\mu}(\tau) &= - \frac{\eta^{\mu\nu}}{2} \left(2\partial_\alpha h_{\nu\beta} - \partial_\nu h_{\alpha\beta}\right)\vert_{\gamma_0(\tau)} u^\alpha u^\beta \,.
\end{align}
Since this curve solves the geodesic equation to first order in epsilon
and we have parametrized it with respect to the particles proper time, 
the on-shell condition
\begin{align}
    1 \equiv g_{\mu\nu}(\gamma(\tau)) \dot{\gamma}^{\mu}(\tau) \dot{\gamma}^{\nu}(\tau)\,,
\end{align}
enforced by the BC, also holds along the whole trajectory.
In our specific scenario the on-shell relation can be approximated as
\begin{align}
    1 \approx (\eta_{\mu\nu} + \epsilon h_{\mu\nu}(\gamma_0(\tau))) u^{\mu}u^{\nu} + \underbrace{2 \eta_{\mu\nu} \Delta \epsilon \dot{\gamma}^{\mu}(\tau) u^{\nu}}_{(\ast)} \,.
\end{align}
The last term, \((\ast)\), expands to
\begin{align}
    2 \eta_{\mu\nu} \Delta \epsilon \dot{\gamma}^{\mu}(\tau) u^{\nu} &= 
    - \frac{2\epsilon}{\tau_D - \tau_P} \int_{\tau_P}^{\tau_D}\int_{\tau_P}^{\tau} \eta_{\mu\nu} \Delta a^{\mu}(\tau^{\prime}) u^{\nu} \;\mathrm{d}\tau^{\prime}\mathrm{d}\tau
    + 2\epsilon\int_{\tau_P}^{\tau} \eta_{\mu\nu} \Delta a^{\mu}(\tau^{\prime}) u^{\nu} \;\mathrm{d}\tau^{\prime} \,,
\end{align}
where
\begin{align}
    2 \eta_{\mu\nu} \Delta a^{\mu}(\tau^{\prime}) u^{\nu} &= -2\eta_{\mu\nu} \frac{\eta^{\mu\sigma}}{2} \left(2\partial_\alpha h_{\sigma\beta} - \partial_\sigma h_{\alpha\beta}\right)\vert_{\gamma_0(\tau)} u^\alpha u^\beta u^{\nu} \\
    &= - \left(2\partial_\alpha h_{\nu\beta} - \partial_\nu h_{\alpha\beta}\right)\vert_{\gamma_0(\tau)} u^\alpha u^\beta u^{\nu} \\
    &= - \partial_\alpha h_{\nu\beta}(\gamma_0(\tau)) u^\alpha u^\beta u^{\nu} \\
    &= - \frac{\mathrm{d}}{\mathrm{d} \tau} \left(h_{\nu\beta}(\gamma_0(\tau)) u^\beta u^{\nu}\right) \,.
\end{align}
Plugging this result back into \((\ast)\) and using the fundamental theorem of calculus yields
\begin{align}
    \begin{split}
        2 \eta_{\mu\nu} \epsilon \Delta \dot{\gamma}^{\mu}(\tau) u^{\nu} &= 
        \frac{\epsilon}{\tau_D - \tau_P} \int_{\tau_P}^{\tau_D} \left[h_{\alpha\beta}(\gamma_0(\tau)) u^{\alpha} u^\beta - h_{\alpha\beta}(X_P) u^{\alpha} u^{\beta} \right] \; \mathrm{d}\tau \\
        &\quad - \epsilon\left[h_{\alpha\beta}(\gamma_0(\tau)) u^{\alpha} u^\beta - h_{\alpha\beta}(X_P) u^{\alpha} u^{\beta} \right]
    \end{split} \\
    &= \frac{\epsilon}{\tau_D - \tau_P} \int_{\tau_P}^{\tau_D} h_{\alpha\beta}(\gamma_0(\tau)) u^{\alpha} u^\beta \; \mathrm{d}\tau
    - \epsilon h_{\alpha\beta}(\gamma_0(\tau)) u^{\alpha} u^\beta\,.
\end{align}
Finally, substituting \((\ast)\) back into the on-shell condition yields
\begin{align}
    1 &\approx (\eta_{\mu\nu} + \epsilon h_{\mu\nu}(\gamma_0(\tau))) u^{\mu}u^{\nu} + \frac{\epsilon}{\tau_D - \tau_P} \int_{\tau_P}^{\tau_D} h_{\alpha\beta}(\gamma_0(\tau)) u^{\alpha} u^\beta \; \mathrm{d}\tau
    - \epsilon h_{\alpha\beta}(\gamma_0(\tau)) u^{\alpha} u^\beta \\
    &= \eta_{\mu\nu} u^{\mu}u^{\nu} + \frac{\epsilon}{\tau_D - \tau_P} \int_{\tau_P}^{\tau_D} h_{\mu\nu}(\gamma_0(\tau)) u^{\mu} u^\nu \; \mathrm{d}\tau \label{eqn:onshell}\,.
\end{align}
In the next section, we will use this result and the ultra relativistic approximation
in order to obtain the correction to the spatial distance traveled by a neutrino in a GW spacetime.

\section{Modified Neutrino Path Length for general Gravitational Waves}
\label{app:ModPath}

Using equation~\eqref{eqn:Length_with_Geodesic} and the \(\mathcal{O}(\epsilon)\) solution to the geodesic equation 
found in the last section, we are now able to compute the corrected physical length.
First we show that \(L(t_P, t_D) \approx t_D - t_P\) in the ultra relativistic limit and subsequently, we use this fact and the on-shell condition~\eqref{eqn:onshell} to derive a relation between the physical, spatial distance between neutrino production and detection and the coordinate distance,
\begin{align}
    L_0 := \vert\vert \vec{X}_D - \vec{X}_P \vert\vert \,.
\end{align}
The latter distance may be identified with the flat spacetime distance, as the physical distance reduces to the coordinate distance for \(h \equiv 0\).
Indeed we will see that in most cases the physical distance oscillates around this flat spacetime value justifying that to some degree it still has physical significance.

First we prove that \(L(t_P, t_D) \approx t_D - t_P\) by evaluating equation~\eqref{eqn:Length_with_Geodesic}.
Since we have already rewritten \(L\) in terms of \(\dot{\gamma^{0}}\) using TT gauge this task is trivial:
\begin{align}
    L(t_P, t_D) = \int_{\tau_P = \tau(t_P)}^{\tau_D = \tau(t_D)} \left(\left(\dot{\gamma}^0\right)^2 - 1\right)^{\frac{1}{2}}\text{d}\tau
    \stackrel{\dot{\gamma}^0 \gg 1}{\approx} \int_{\tau_P}^{\tau_D} \dot{\gamma}^0 \text{d}\tau
    = t_D - t_P \,.
\end{align}
The other important ingredient is the deviation of the flow of proper time from the flat spacetime case quantified in equation~\eqref{eqn:onshell},
\begin{align}
    1 &= \eta_{\mu\nu} u^{\mu}u^{\nu} + \frac{\epsilon}{\tau_D - \tau_P} \int_{\tau_P}^{\tau_D} h_{\mu\nu}(\gamma_0(\tau)) u^{\mu} u^\nu \; \mathrm{d}\tau \\
    \Leftrightarrow (u^0)^2 &\approx \vec{u}^2 - \frac{\epsilon}{\tau_D - \tau_P} \int_{\tau_P}^{\tau_D} h_{\mu\nu}(\gamma_0(\tau)) u^{\mu} u^\nu \; \mathrm{d}\tau \\
    \Leftrightarrow \frac{(t_D - t_P)^2}{(\tau_D - \tau_P)^2} &\approx \frac{\vert\vert \vec{X}_D - \vec{X}_P \vert\vert^2}{(\tau_D - \tau_P)^2} - \frac{\epsilon}{\tau_D - \tau_P} \int_{\tau_P}^{\tau_D} h_{\mu\nu}(\gamma_0(\tau)) u^{\mu} u^\nu \; \mathrm{d}\tau \\
    \Leftrightarrow t_D - t_P &\approx \left[L_0^2 - \epsilon (\tau_D - \tau_P) \int_{\tau_P}^{\tau_D} h_{\mu\nu}(\gamma_0(\tau)) u^{\mu} u^\nu \; \mathrm{d}\tau \right]^{\frac{1}{2}} \\
    \Leftrightarrow t_D - t_P &\approx L_0\left[1 - \epsilon (\tau_D - \tau_P) \frac{u^j}{L_0} \frac{u^k}{L_0} \int_{\tau_P}^{\tau_D} h_{jk}(\gamma_0(\tau)) \; \mathrm{d}\tau \right]^{\frac{1}{2}} \\
    \Leftrightarrow t_D - t_P &\approx L_0 - \frac{\epsilon}{2} L_0 (\tau_D - \tau_P) \frac{u^j}{L_0} \frac{u^k}{L_0} \int_{\tau_P}^{\tau_D} h_{jk}(\gamma_0(\tau)) \; \mathrm{d}\tau \,.
\end{align}
Here, we used that \(h\) is given in the TT gauge in order to eliminate all terms involving \(h_{0\mu}\).
Next, we exploit \(L \approx t_D - t_P\) and \(u^j = (X_D^j - X_P^j) / (\tau_D - \tau_P)\) and define the unit vectors 
\(\hat{l}^j := (X_D^j - X_P^j) / L_0\) pointing into the direction of neutrino flow.
Then we can write the physical length as
\begin{align}
    L(t_P, t_D) \approx L_0 - \frac{\epsilon}{2} \frac{L_0}{\tau_D - \tau_P} \hat{l}^j \hat{l}^k 
    \int_{\tau_P}^{\tau_D} h_{jk}(\gamma_0(\tau)) \;\mathrm{d}\tau \,.
\end{align}
Furthermore neglecting terms of \(\mathcal{O}(\epsilon^2)\), we can write \(\frac{L_0}{\tau_D - \tau_P} \approx u^0\).
Thus the most general expression without further specifying \(h_{\mu\nu}\) reads
\begin{align}
    L(t_P, t_D) &\approx L_0 - \frac{\epsilon}{2} u^0 
    \int_{\tau_P}^{\tau_D} h_{\parallel}(\gamma_0(\tau)) \;\mathrm{d}\tau \,, \\
    h_{\parallel}(\gamma_0(\tau)) &:= \hat{l}^j \hat{l}^k h_{jk}(\gamma_0(\tau)) \,.
\end{align}
We can make this formula more explicit by employing the Fourier decomposition of \(h_{\mu\nu}\) in plane wave modes which is given by
\begin{align}
    h_{\mu\nu}(x) &= \sum_{r \in \{+, \times\}} \int \mathrm{d}^3\vec{k} \; \mathrm{Re}\left[\Psi_r(\vec{k}) A_{\mu\nu}^{r}(\hat{k}) \exp(-i k x)\right] \,,
    \label{eqn:generalGW}
\end{align}
where \(\Psi_r\) are the complex, scalar weight function (wave packets) of the corresponding modes and \(A_{\mu\nu}^{r}(\hat{k})\)
are the normalized polarization tensors depending only on the direction \(\hat{k}\) of the respective mode.
Next we briefly discuss the polarization tensors and our choice of coordinates before we carry on the calculation
of the integrated GW tensor along the neutrino trajectory.

A reasonable choice of coordinate system to describe the neutrino oscillation experiment of interest 
would be to define the coordinate axes such that the neutrino source resides at \(\vec{X}_P \equiv 0\) and
the detector remains at \(\vec{X}_D \equiv L_0 \vec{e}_z\).
This way we also assume that source and detector are not bound to each other by any forces since the physical length can vary freely
as the GW passes through the system.
Furthermore, we define the propagation direction of the GW with respect to this coordinate system.
In order to obtain the parallel polarization components
\begin{align}
    A_{\parallel}^{r}(\hat{k}) &:= \hat{l}^j \hat{l}^k A_{lk}^{r}(\hat{k}) \,,
\end{align}
which we need for our length expression, we simply rotate from the coordinate system, \(\tilde{\Sigma}\), 
where \(\hat{k} = \vec{e}_z\) to our system, \(\Sigma\), using one rotation around the \(y\) axis with 
angle \(\theta\) followed by a rotation around the \(z\) axis with angle \(\varphi\).
This yields
\begin{align}
    A_{\parallel}^{r}(\hat{k}) &= \hat{l}^j \hat{l}^k A_{jk}^{r}(\hat{k}) \\
    &= R^{m}{}_{j} \hat{l}^j R^{n}{}_{k} \hat{l}^k \tilde{A}_{mn}^{r}(R^T \hat{k}) \\
    &= R^{m}{}_{3} R^{n}{}_{3} \tilde{A}_{mn}^{r}(\hat{e}_z) \\
    &= \begin{pmatrix} \cos(\varphi) \sin(\theta) \\ \sin(\varphi) \sin(\theta) \\ \cos(\theta) \end{pmatrix}^T
    \begin{pmatrix}
        A_{11}^r    & A_{12}^r  & 0 \\
        A_{12}^r    & -A_{11}^r & 0 \\
        0           & 0         & 0 \\
    \end{pmatrix}
    \begin{pmatrix} \cos(\varphi) \sin(\theta) \\ \sin(\varphi) \sin(\theta) \\ \cos(\theta) \end{pmatrix} \,,
\end{align}
where \(A^{+}_{11} = 1\), \(A^{+}_{12} = 0\) and \(A^{\times}_{11} = 0\), \(A^{\times}_{12} = 1\).
Multiplying everything out, we get
\begin{align}
    A_{\parallel}^{+}(\hat{k}) &= \sin^2(\theta) (\cos^2(\varphi) - \sin^2(\varphi)) = \sin^2(\theta) \cos(2\varphi) \,, \\
    A_{\parallel}^{\times}(\hat{k}) &= 2 \sin^2(\theta) \cos(\varphi) \sin(\varphi) = \sin^2(\theta) \sin(2\varphi) \,.
\end{align}
The angle \(\theta\) describes the angle between the zeroth order neutrino trajectory and the GW propagation direction while
\(\varphi\) quantifies the direction of the GW perpendicular to the neutrino trajectory.

Next, we decompose the complex wave functions \(\Psi_r\) into their absolute values and phases, i.e.
\begin{align}
    \Psi_r(\vec{k}) =: \psi_r(\vec{k}) \exp(-i \phi^r(\vec{k})) \,,
\end{align}
leading us to the final expression for \(h_{\parallel}\):
\begin{align}
    h_{\parallel}(x) &= \sum_{r \in \{+, \times\}} \int \mathrm{d}^3\vec{k} \; \psi_r(\vec{k}) A_{\parallel}^{r}(\varphi, \theta) \cos(k x + \phi^r(\vec{k}))\,.
\end{align}
Using this result, we can perform the integration along \(\gamma_0\) explicitly and obtain a formula for \(\Delta L\) in terms of the momentum distribution of \(h\):
\begin{align}
    \Delta L(t_P, t_D) &= - \frac{\epsilon}{2} u^0 \int_{\tau_P}^{\tau_D} h_{\parallel}(\gamma_0(\tau)) \; \mathrm{d}\tau \\
    &= - \frac{\epsilon}{2} u^0 \sum_{r \in \{+, \times\}} \int \mathrm{d}^3\vec{k} \; 
    \psi_r(\vec{k}) A_{\parallel}^{r}(\varphi, \theta)
    \int_{\tau_P}^{\tau_D} \cos(k \gamma^0 + \phi^r(\vec{k})) \; \mathrm{d}\tau \\
    &= - \frac{\epsilon}{2} \sum_{r \in \{+, \times\}} \int \mathrm{d}^3\vec{k} \; 
    \frac{u^0}{k u} \psi_r(\vec{k}) A_{\parallel}^{r}(\varphi, \theta)
    \int_{k \gamma_0(\tau_P) + \phi^r(\vec{k})}^{k \gamma_0(\tau_D) + \phi^r(\vec{k})} \cos(\xi) \; \mathrm{d}\xi \\
    &\approx - \frac{\epsilon}{2} \sum_{r \in \{+, \times\}} \int \mathrm{d}^3\vec{k} \; 
    \frac{\psi_r(\vec{k}) A_{\parallel}^{r}(\varphi, \theta)}{\omega (1 - \cos(\theta))}
    \int_{k \gamma_0(\tau_P) + \phi^r(\vec{k})}^{k \gamma_0(\tau_D) + \phi^r(\vec{k})} \cos(\xi) \; \mathrm{d}\xi \\
    &= - \frac{\epsilon}{2} \sum_{r \in \{+, \times\}} \int \mathrm{d}^3\vec{k} \; 
    \frac{\psi_r(\vec{k}) A_{\parallel}^{r}(\varphi, \theta)}{\omega (1 - \cos(\theta))}
    \left(\sin(k X_D + \phi^r(\vec{k})) - \sin(k X_P + \phi^r(\vec{k}))\right) \,,
\end{align}
where \(\omega = \vert\vert \vec{k} \vert\vert\) and \(\vec{u}\vec{k} \approx u^0 \omega \cos(\theta)\).

Again making use of our choice of coordinates, the ultrarelativistic limit and neglecting \(\mathcal{O}(\epsilon^2)\) terms, we can write \(X_P = (t, \vec{0})\) and \(X_D = (t + L_0, L_0 \vec{e}_z)\) for some time \(t\) parameterizing the trajectories of neutrino source and detector.
Thus, from now on we will also parametrize the distance deviation in terms of \(t\), i.e.
\begin{align}
    \Delta L(t) \approx - \frac{\epsilon}{2} \sum_{r \in \{+, \times\}} \int \mathrm{d}^3\vec{k} \; 
    \frac{\psi_r(\vec{k}) A_{\parallel}^{r}(\varphi, \theta)}{\omega (1 - \cos(\theta))}
    &\left(\sin(\omega (t + L_0) - \omega L_0 \cos(\theta) + \phi^r(\vec{k})) - \sin(\omega t + \phi^r(\vec{k}))\right) \\
    \begin{split}
        = - \frac{\epsilon}{2} \sum_{r \in \{+, \times\}} \int \mathrm{d}^3\vec{k} \; 
        \frac{\psi_r(\vec{k}) A_{\parallel}^{r}(\varphi, \theta)}{\omega (1 - \cos(\theta))}
        &\left(\sin\left\{\omega (1 - \cos(\theta) )L_0\right\} \cos(\omega t + \phi^r(\vec{k})) \right.\\
        &\quad + \left. \left[\cos\left\{\omega (1 - \cos(\theta) )L_0\right\} - 1\right] \sin(\omega t + \phi^r(\vec{k}))\right) \,.
    \end{split}
\end{align}
Lastly, we want to consider the case of a sharply peaked momentum space wave packet, i.e.
\begin{align}
    \psi_r(\vec{k}) &= a_r \delta_{\varepsilon}^{(3)}(\vec{k} - \vec{K}) \,,
\end{align}
where \(\delta_{\varepsilon}^{(3)}\) represents a sequence of functions converging to the delta distribution under the integral sign for \(\varepsilon \to 0\).
In the case where all other functions than \(\psi_r\) are sufficiently weakly momentum dependent on the support of \(\delta_{\varepsilon}\)
we can pull them out of the integral and evaluate them at the peak momentum \(\vec{K}\) which yields
\begin{multline}
    \Delta L(t) = - \frac{\epsilon}{2} \sum_{r \in \{+, \times\}} 
    \frac{a_r A_{\parallel}^{r}(\varphi, \theta)}{\omega_{K} (1 - \cos(\theta))}
    \left(\sin\left\{\omega_{K} (1 - \cos(\theta) )L_0\right\} \cos(\omega_{K} t + \phi^r(\vec{K})) \right.\\
    \quad + \left. \left[\cos\left\{\omega_K (1 - \cos(\theta) )L_0\right\} - 1\right] \sin(\omega_K t + \phi^r(\vec{K}))\right) \,.
\end{multline}

\section{The influence of non-trivial Neutrino Trajectories on the Gravitational Wave Effect}
\label{app:EarthRotation}
Especially, when considering neutrinos in the solar system, the orientation of the GW and neutrino trajectories
might vary non-trivially during the time of data taking.
Usually this includes the motion of the neutrino detector and source relatively to each other and the orbit
of their center of mass around the sun.
Consequently this also has an non-negligible impact on the effect of the GW on the neutrino flavor transition probability
since the angle between the GW direction \(\vec{k}\) and the zeroth order neutrino trajectory direction \(\vec{v}\) might become time dependent.
Figure~\ref{fig:Earth_Rotation} illustrates the scenario just discussed.
\begin{figure}
  \centering
      \begin{tikzpicture}
          \draw[fill=orange] (0,0) circle [radius=2];
          \node[below] at (0,0) {Sun};
          \draw[even odd rule]
            (0,-.2) circle [x radius=6, y radius=1];
          \begin{scope}
            \clip[overlay] (-2.1,0) rectangle (2.1,2.1);
            \draw[fill=orange] (0,0) circle [radius=2];
          \end{scope}
          \draw (6,0) circle [radius=1];
          \draw [black] (5,-2) -- (7,2);
          \node [right, black] at (7,2) {Rotation axis};
          \draw [dashed, white] (6,0) circle [radius = 1.25];
          \draw [fill=red] (7, 0) circle [radius=0.1];
          \node[right, red] at (7, 0) {\small Production};
          \draw [fill=gray] (4.9, 0) circle [radius=0.1];
          \node [left, gray] at (4.9, 0) {\small Detection};
          \draw [dashed] (4.9, 0) -- (7, 0);
          \draw [dashed, gray] (7,-2) -- (6.5,-1.5);
          \draw [dashed, gray] (8,-2) -- (7,-1);
          \draw [dashed, gray] (9,-2) -- (8.45,-1.45);
          \draw [dashed, gray] (7.75,0.25) -- (6.75,1.25);
          \draw [dashed, gray] (9,0) -- (7.5,1.5);
          \draw [dashed, gray] (4.25,0.75) -- (3,2);
          \draw [dashed, gray] (5,1) -- (4,2);
          \draw [dashed, gray] (5.5,1.5) -- (5,2);
          \draw [dashed, gray] (6.5,1.5) -- (6,2);
          \draw [dashed, gray] (6,-2) -- (5.5,-1.5);
          \draw [dashed, gray] (3.25,0.75) -- (2,2);
          \draw [->, gray] (9,-0.75) -- (8, -1.75); 
          \draw [->, black] (9,-0.75) -- (7.75, -0.75); 
          \node [right, gray] at (8.5, -1.3) {\small $\vec{k}_{GW}$};
          \node [above, black] at (8.37, -0.75) {\small $\vec{k}_{\nu}$};
          \coordinate (A) at (8.7,-0.75);
          \coordinate (B) at (9,-0.75);
          \coordinate (C) at (8.7, -1.05);
          \pic [draw, <->, angle radius=9.5mm] {angle = A--B--C};
          \node [black, right] at (8,-1) {\rotatebox{22.5}{\footnotesize $\omega t
          $}};
        \end{tikzpicture}
  \caption{Illustration of the rotation of the detction and production region around the sun. The GW background is shown by the dashed gray lines with a resulting angle $\Theta (t)$ between the direction of propagation of the GW $\vec{k}_\mathrm{GW}$ and the neutrino propagation $\vec{k}_\nu$.}
  \label{fig:Earth_Rotation}
\end{figure}
In the following, we assume that the relative motion of neutrino source and detector is negligible during the propagation time of a single neutrino.
This will later allow us to repeat most of the steps of the derivation of \(\Delta L\) shown in appendix~\ref{app:ModPath} 
since during the integration over the neutrino path we can neglect the newly introduced time dependence.
On the time scale of the neutrino oscillation experiment, however, the relative orientation of the neutrino velocity and GW vector 
as well as the GW tensor components may change significantly.

Assuming that the GW direction remains fixed in the sun's rest frame, \(\tilde{\Sigma}\) described by coordinates \(\tilde{x}^{\mu}\),
enables us to calculate the now time dependent GW parameters in the co-rotating center of mass system (CMS) of the neutrino source-detector system, \(\Sigma\),
via a coordinate change to the new coordinates \(x^{\mu}(\tilde{x}^{\alpha})\).
The corresponding GW tensors are, of course, related by the usual transformation law
\begin{align}
  h_{\alpha\beta}(x) &= \frac{\partial \tilde{x}^{\mu}}{\partial x^{\alpha}}\frac{\partial \tilde{x}^{\nu}}{\partial x^{\beta}}
  \tilde{h}_{\mu\nu}(\tilde{x}(x))\,,
\end{align}
where the GW tensor in the sun's rest frame, \(\tilde{h}_{\mu\nu}\), is still described by
\begin{align}
  \tilde{h}_{\mu\nu}(\tilde{x}) &= \sum_{r \in \{+, \times\}} \int \mathrm{d}^3\vec{k} \; 
  \psi_r(\vec{k}) A_{\mu\nu}^{r}(\hat{k}) \cos(k \tilde{x} + \phi^r(\vec{k})) \,,
  \label{eqn:GW_Sun_Restframe}
\end{align}
c.f. equation~\eqref{eqn:generalGW}, where we take \(\Psi_r = \psi_r \exp(-i \phi^r)\) with \(\psi_r \geq 0\).

In order to find the GW tensor in the CMS of the neutrino experiment, our task now is to derive an expression for the coordinate transformation \(\tilde{x}^{\mu}(x^{\alpha})\).
To that end, assume for simplicity that neutrino source and detector move at a constant coordinate distance \(L_0\), 
as described in appendix~\ref{app:ModPath}.
Now, let \(X_P(t)\) and \(X_D(t)\) be the trajectories of corresponding neutrino production and detection events, respectively.
We define absolute and relative coordinates in \(\tilde{\Sigma}\) by the relations
\begin{align}
  \begin{rcases}
    \tilde{D}^{\mu}(t) := \frac{\tilde{X}_P^{\mu}(t) + \tilde{X}_D^{\mu}(t)}{2} \\
    \tilde{r}^{\mu}(t) := \frac{\tilde{X}_D^{\mu}(t) - \tilde{X}_P^{\mu}(t)}{2}
  \end{rcases} \Leftrightarrow
  \begin{cases}
    \tilde{X}_P^{\mu}(t) := \tilde{D}^{\mu}(t) - \tilde{r}^{\mu}(t) \\
    \tilde{X}_D^{\mu}(t) := \tilde{D}^{\mu}(t) + \tilde{r}^{\mu}(t) \\
  \end{cases} \,,
\end{align}
where \(\vert\vert \vec{\tilde{r}}(t) \vert\vert \equiv L_0 / 2\) and \(\tilde{r}^{0}(t) \equiv L_0 / 2\) at zeroth order in the GW and using the ultrarelativistic approximation.
Thus, at most \(\tilde{r}\) performs a rotating motion in \(\tilde{\Sigma}\).
We specify the coordinate system \(\Sigma\) by demanding that
\begin{itemize}
  \item \(t = \tilde{t}\)
  \item \(\vec{D}(t) \equiv 0\)
  \item \(\vec{r}(t) \equiv \frac{L_0}{2} \vec{e}_z\)
\end{itemize}
leading to the transformation law
\begin{align}
  \begin{rcases}
    t(\tilde{x}) = t \\
    \vec{x}(\tilde{x}) = R^T(\vec{\Phi}) R^T(\vec{\Theta}(t))\left(\vec{\tilde{x}} - \vec{\tilde{D}}(\tilde{t})\right)
  \end{rcases} \Leftrightarrow
  \begin{cases}
    \tilde{t}(x) = t \\
    \vec{\tilde{x}}(x) = \vec{\tilde{D}}(t) + R(\vec{\Theta}(t)) R(\hat{\Phi}) \vec{x}
  \end{cases} \,,
\end{align}
where \(\Theta(t)\) is the axis (and angle) of rotation about which \(\vec{\tilde{r}}\) rotates in \(\tilde{\Sigma}\)
and \(\vec{\Phi}\) is the rotation axis (and angle) needed to rotate \(L_0 \cdot \hat{\tilde{e}}_z / 2\) to \(\vec{\tilde{r}}(0)\).
From now on, we only refer to this combination of rotations as \(O(t) := R(\vec{\Theta}(t)) R(\vec{\Phi})\).
The components of the Jacobian of this coordinate transformation thus read
\begin{align}
  \frac{\partial \tilde{x}^{0}}{\partial x^{0}} &= 1 \,, 
  &&\frac{\partial \tilde{x}^{0}}{\partial x^{k}} = 0 \,,
  &&\frac{\partial \tilde{x}^{j}}{\partial x^{0}} = \dot{\tilde{D}}^j(t) + \dot{O}^{j}{}_{l}(t) x^l \,,
  &&\frac{\partial \tilde{x}^{j}}{\partial x^{k}} = O^{j}{}_{k}(t) \,.
\end{align}
And the GW tensor becomes
\begin{align}
  \begin{split}
    h_{\alpha\beta}(\tilde{x}) &= \left(O^{\mu}{}_{\alpha}(t) + \delta^{\mu}(t) \delta_{\alpha 0}\right)
    \left(O^{\nu}{}_{\beta}(t) + \delta^{\nu}(t) \delta_{\beta 0}\right) \\
    & \quad \times \sum_{r \in \{+, \times\}} \int \mathrm{d}^3\vec{k} \; \psi_r(\vec{k}) A_{\mu\nu}^{r}(\hat{k}) 
    \cos\left(\omega t - \vec{k}^T O^T(t) (\vec{x} - \vec{\tilde{D}}(t)) + \phi^r(\vec{k})\right)
  \end{split} \\
  \begin{split}
    &= \sum_{r \in \{+, \times\}} \int \mathrm{d}^3\vec{k} \; \psi_r(\vec{k}) 
    O^{\mu}{}_{\alpha}(t) O^{\nu}{}_{\beta}(t) A_{\mu\nu}^{r}(\hat{k}) 
    \cos\left(\omega t - (O(t)\vec{k})^T \vec{x} + (O(t)\vec{k})^T \vec{\tilde{D}}(t) + \phi^r(\vec{k})\right) \\
    &\quad + \delta_{\alpha 0} \sum_{r \in \{+, \times\}} \int \mathrm{d}^3\vec{k} \; \psi_r(\vec{k}) 
    \delta^{\mu}(t) O^{\nu}{}_{\beta}(t) A_{\mu\nu}^{r}(\hat{k}) 
    \cos\left(\omega t - (O(t)\vec{k})^T \vec{x} + (O(t)\vec{k})^T \vec{\tilde{D}}(t) + \phi^r(\vec{k})\right) \\
    &\quad + \delta_{\beta 0} \sum_{r \in \{+, \times\}} \int \mathrm{d}^3\vec{k} \; \psi_r(\vec{k}) 
    O^{\mu}{}_{\alpha}(t) \delta^{\nu}(t) A_{\mu\nu}^{r}(\hat{k}) 
    \cos\left(\omega t - (O(t)\vec{k})^T \vec{x} + (O(t)\vec{k})^T \vec{\tilde{D}}(t) + \phi^r(\vec{k})\right) \\
    &\quad + \delta_{\alpha 0} \delta_{\beta 0} \sum_{r \in \{+, \times\}} \int \mathrm{d}^3\vec{k} \; \psi_r(\vec{k}) 
    \delta^{\mu}(t) \delta^{\nu}(t) A_{\mu\nu}^{r}(\hat{k}) 
    \cos\left(\omega t - (O(t)\vec{k})^T \vec{x} + (O(t)\vec{k})^T \vec{\tilde{D}}(t) + \phi^r(\vec{k})\right) \,,
  \end{split}
\end{align}
where we define \(\delta^j = \dot{\tilde{D}}^j(t) + \dot{O}^{j}{}_{l}(t) x^l\) and \(\delta^{0} = 0\).
To simplify this expression, we perform an orthogonal transformation of the integration variable \(\vec{q} = O(t) \vec{k}\)
which we then rename to \(\vec{k}\) again.
Furthermore, we exploit the transformation behavior of the polarization tensor 
\(A_{\mu\nu}(R \hat{k}) = {R}^{\alpha}{}_{\mu}{R}^{\beta}{}_{\nu} A_{\alpha\beta}(\hat{k})\) and define 
\(\chi^r(\vec{k}, t) := \phi^r(\vec{k}) + \vec{k}\vec{\tilde{D}}(t)\).
This yields
\begin{align}
  \begin{split}
    h_{\alpha\beta}(\tilde{x}) &= \sum_{r \in \{+, \times\}} \int \mathrm{d}^3\vec{k} \; \psi_r(O^T \vec{k}) A_{\alpha\beta}^{r}(\hat{k}) 
    \cos\left(\omega t - \vec{k} \cdot \vec{x} + \chi^r(\vec{k}, t)\right) \\
    &\quad + \delta_{\alpha 0} \sum_{r \in \{+, \times\}} \int \mathrm{d}^3\vec{k} \; \psi_r(\vec{k}) 
    \delta^{\mu}(t) {O^T}^{\rho}{}_{\mu} A_{\rho\beta}^{r}(\hat{k}) 
    \cos\left(\omega t - \vec{k} \cdot \vec{x} + \chi^r(\vec{k}, t)\right) \\
    &\quad + \delta_{\beta 0} \sum_{r \in \{+, \times\}} \int \mathrm{d}^3\vec{k} \; \psi_r(\vec{k}) 
    \delta^{\nu}(t) {O^T}^{\sigma}{}_{\nu} A_{\alpha\sigma}^{r}(\hat{k}) 
    \cos\left(\omega t - \vec{k} \cdot \vec{x} + \chi^r(\vec{k}, t)\right) \\
    &\quad + \delta_{\alpha 0} \delta_{\beta 0} \sum_{r \in \{+, \times\}} \int \mathrm{d}^3\vec{k} \; \psi_r(\vec{k}) 
    \delta^{\mu}(t) {O^T}^{\rho}{}_{\mu} \delta^{\nu}(t) {O^T}^{\sigma}{}_{\nu} A_{\sigma\rho}^{r}(\hat{k}) 
    \cos\left(\omega t - \vec{k} \cdot \vec{x} + \chi^r(\vec{k}, t)\right) \,.
  \end{split}
\end{align}
One can immediately notice that this coordinate system does not belong to the TT gauge orbit since \(h_{0\mu} \neq 0\) in general.
Thus in the most general case we get corrections to \(\Delta L\) from the non-inertial coordinate transformation we just discussed.
However, when considering sufficiently slow motion of the coordinate system during the propagation of a single neutrino, i.e. 
\begin{align}
  (t_D - t_P) \cdot \vert \dot{\delta}^j(t) \vert \ll 1 \,,
\end{align}
we can actually neglect these terms in our expression in \(\Delta L\) as being of higher than linear order in small quantities.
Thus in this approximation the TT gauge condition holds again and we can proceed to calculate \(\Delta L\) as before in appendix~\ref{app:ModPath}, yielding
\begin{align}
  \Delta L(t) &\approx - \frac{\epsilon}{2} \sum_{r \in \{+, \times\}} \int \mathrm{d}^3\vec{k} \; 
  \frac{\psi_r(O^T(t) \vec{k}) A_{\parallel}^{r}(\hat{k})}{\omega - \vec{k}\vec{v}}
  \left(\sin(k X_D(t) + \chi^r(\vec{k}, t)) - \sin(k X_P(t) + \chi^r(\vec{k}))\right) \,.
\end{align}
In order to obtain a better impression on what changes compared to the case discussed in appendix~\ref{app:ModPath},
we now apply the inverse integral transform as before leading to
\begin{align}
  \Delta L(t) &\approx - \frac{\epsilon}{2} \sum_{r \in \{+, \times\}} \int \mathrm{d}^3\vec{k} \; 
  \frac{\psi_r(\vec{k}) A_{\parallel}^{r}(O(t)\hat{k})}{\omega - \vec{k}O(t)\vec{v}}
  \left(\sin(k O(t)X_D(t) + \chi^{\prime r}(\vec{k}, t)) - \sin(k O(t)X_P(t) + \chi^{\prime r}(\vec{k}))\right) \,.
\end{align}
Therefore, we see that the relative orientation of the neutrino trajectory and the GW direction now changes with time, as expected,
implying a variable influence of the effect.
If for example \(\vec{k}\) becomes (anti-)parallel to \(\vec{v}\) the polarization vector vanishes and the corresponding mode doesn't contribute anymore.


\section{Toy Analysis: Likelihoods}
\label{app:Statistics}

In this section, we discuss the statistical procedure used to obtain the exclusion limits for the GW parameter space.
First, we describe the setup of a toy experiment roughly modeling a typical neutrino oscillation experiment.
Afterwards, we describe likelihood ratio tests and how we use them in order to obtain the allowed parameter space of our model for a given toy data set generated assuming no GW background.

\subsection{Experimental Data Model}

We attempt to model the unfolded experimental neutrino event counts for each flavor. To keep things simple since this is 
meant to be a proof of principle analysis rather than a complete experimental treatment, we neglect the following effects: 
\begin{itemize}
    \item Matter effects
    \item Systematic uncertainties like detector efficiency, accuracy and acceptance
    \item The procedure of unfolding the measured data
    \item Theoretical uncertainties
\end{itemize}
Furthermore, we do not simulate multiple data sets by sampling random event counts from the corresponding Poisson distribution.
Instead, we take the underlying theory prediction for the expected number of events as the outcome of our experiment. 
The details of the statistical methods employed in this work will be described in the next subsection.

Regarding the experimental setup, we take into account the finite energy resolution and the binning of neutrino counts. 
Moreover, experiments never directly measure the flavor transition probabilities, but only in convolution with the energy flux spectrum.
The form of this spectrum mostly depends on the origin of neutrinos, e.g. an energy power-law for atmospheric and astrophysical neutrinos.

Therefore, we need to compute the expected number of events from the combined probability distribution function (pdf):
\begin{equation}
    \rho_b (E, L) = \sum_a \varphi_a(E) P_{a b} (E, L_0)\,,
\end{equation}
where $\varphi_a$ is the normalized neutrino flux for an initial flavor $a$ and $P_{ab}$ is the transition 
probability of a neutrino from flavor $a$ to flavor $b$.
Since we aim to predict the event count for each energy bin, we need the accumulated probability for the respective energy ranges.
Thus, for the $i$-th energy bin and flavor $b$ we arrive at the final discrete probabilities:
\begin{equation}
    p_{i b} (L_0) := \int_{E_i}^{E_{i+1}} \rho_b (E, L_0) \; \text{d} E \,,
\end{equation}
where $E_i$ and $E_{i+1}$ are the lower and upper bin edges of the $i$-th bin as illustrated in figure~\ref{fig:Likelihood}.
\definecolor{LimeGreen}{HTML}{A5DB72}
\definecolor{Fuchsia}{HTML}{DB7D60}
\definecolor{Blueby}{HTML}{5D75D9}
\begin{figure}
    \centering
    \begin{tikzpicture}
        \draw[->] (0,0) -- (10,0);
        \draw[->] (0,0) -- (0,5);
        \coordinate[label=right:$E$] (E) at (10,0);
        \coordinate[label=above: $\text{Events}$] (n) at (0,5);
        \fill [fill = LimeGreen] (1,0) rectangle (2,2); 
        \fill [fill = Fuchsia] (1,2) rectangle (2,3);
        \fill [fill = Blueby] (1,3) rectangle (2,4.5);
        \node at (1.5,1) {$\nu_e$};
        \node at (1.5,2.5) {$\nu_\mu$};
        \node at (1.5,3.75) {$\nu_\tau$};
        \draw [thick] (1,-0.25) -- (1,0.25);
        \draw [thick] (2,-0.25) -- (2,0.25);
        \node[below] at (1,-0.25) {$E_1$};
        \node[below] at (2,-0.25) {$E_2$};
        \draw[thick, black, decorate, decoration={brace, mirror, amplitude=10pt}] (1,-1) -- (2,-1) node[midway, below, yshift=-12pt,]{$\text{First energy bin}$};
        \filldraw (3,2.5) circle (2pt);
        \filldraw (4,2.5) circle (2pt);
        \filldraw (5,2.5) circle (2pt);
        \fill [fill = LimeGreen] (6,0) rectangle (7,2.5); 
        \fill [fill = Fuchsia] (6,2.5) rectangle (7,3.2);
        \fill [fill = Blueby] (6,3.2) rectangle (7,4);
        \node at (6.5,1.25) {$\nu_e$};
        \node at (6.5,2.85) {$\nu_\mu$};
        \node at (6.5,3.6) {$\nu_\tau$};
        \draw [thick] (6,-0.25) -- (6,0.25);
        \draw [thick] (7,-0.25) -- (7,0.25);
        \node[below] at (6,-0.25) {$E_n$};
        \node[below] at (7,-0.25) {$E_{n+1}$};
        \draw[thick, black, decorate, decoration={brace, mirror, amplitude=10pt}] (6,-1) -- (7,-1) node[midway, below, yshift=-12pt,]{$n\text{-th energy bin}$};
    \end{tikzpicture}
    \caption{Illustration of the number of events for each flavor in the different energy bins.}
    \label{fig:Likelihood}
\end{figure}

Finally, we want to turn towards modeling the neutrino flux spectrum.
We will especially discuss the neutrino flux expected from natural neutrino sources, i.e. atmospheric and astrophysical neutrinos, and that expected from fixed target experiments.
As already mentioned, in the simplest scenario a naturally produced flux spectrum can be approximated by a power law energy dependence, i.e.
\begin{equation}
    \varphi_a (E) = \frac{c_a}{\phi} E^{-\gamma}\,,
\end{equation}
where $\gamma > 1$, $c_a$ determines the flavor composition of the source flux 
and $\phi$ is a normalization constant such that
\begin{equation}
    1 = \sum_a \int_{E_\text{min}}^{E_\text{max}} \varphi_a(E) \; \text{d} E\,.
\end{equation}
Here, $E_\text{min}$ and $E_\text{max}$ are the minimum and maximum energies of neutrinos emerging from the source, respectively.
Hence, in total we arrive at
\begin{equation}
    \varphi_{a}(E) = (\gamma - 1) \frac{c_{a}}{\sum_{b} c_{b}}
    \frac{E^{-\gamma}}{\left(E_\text{min}^{1-\gamma} - E_\text{max}^{1-\gamma}\right)} \,.
\end{equation}
A typical flavor composition of a neutrino source based on pion decay would for example be $\vec{c} = (c_a)^3_{a = 1} = (1, 2, 0)$.

For a fixed target experiment the precise shape of the flux depends on the details of the neutrino production 
mechanism.
In order to keep our considerations simple, we do not specify a particular production process but approximate the neutrino energy spectrum by a Gaussian shape in order to account for the fact that artificially produced
neutrino spectra usually peak at some mean energy \(\bar{E}\) and exhibit some width \(\sigma_E\).
We adopt this rather crude approximation, since our effect is not sensitive to the exact shape of the energy spectrum.
It is rather the location of the energy peak determining if an effect is visible.
Therefore, in the currently considered setting the neutrino flux spectrum reads
\begin{align}
    \varphi_a(E) &= \frac{c_a}{\phi} \exp{\left(-\frac{1}{2}\left(\frac{E - \bar{E}_a}{\sigma_{E,a}^2}\right)^2\right)} \,,
\end{align}
where the normalization constant is given by
\begin{align}
    \phi &= \sum_{a} c_a \sigma_{E,a} \sqrt{\frac{\pi}{2}} \left(\mathrm{erf}\left(\frac{E_{\mathrm{max}} - \bar{E}_{a}}{\sqrt{2}\sigma_{E, a}}\right)
    - \mathrm{erf}\left(\frac{E_{\mathrm{min}} - \bar{E}_{a}}{\sqrt{2}\sigma_{E, a}}\right)\right) \,.
\end{align}

\subsection{Likelihood Ratio Tests}

In the following, we describe likelihood ratio tests for the example of a coherent GW signal. The procedure 
employed for the SGWB is equivalent by exchanging $h \leftrightarrow A_*$ and $f \leftrightarrow \gamma$.
In order to tell which parameter configurations \((h, \lambda)\) or \((h, f)\), i.e. GW strain and wave length or frequency, 
lead to observable differences in comparison to the scenario where no GW is present, we use a likelihood ratio test.
This method is especially suitable to tell differences between nested theoretical models, i.e. models belonging to 
the same class differing only in their respective parameter configurations.
In our case, the \textit{standard} scenario is represented by the parameter configuration \(h = 0\).

In general the \textit{test statistic} for this test is the \textit{negative logarithmic likelihood ratio} (NLLR) given by
\begin{align}
    \Lambda(\vec{X}) &= -2\ln\left(\frac{\sup_{\vec{\vartheta} \in \Theta_{0}}\mathcal{L}(\vec{\vartheta} \; \vert \; \vec{X})}
    {\sup_{\vec{\vartheta} \in \Theta}\mathcal{L}(\vec{\vartheta} \; \vert \; \vec{X})}\right) \in [0, \infty)\,,
\end{align}
where \(\vec{X}\) is a set of observations, \(\vec{\vartheta}\) is the vector of parameters, \(\Theta_0\) is 
the space of parameter configurations corresponding to the \textit{null hypothesis} (\(H_0\)) and \(\Theta\) represents the space
of all possible parameter configurations.
If \textit{Wilk's theorem} applies, \(\Lambda\) follows a \(\chi^2\) distribution with \(d := \mathrm{dim}(\Theta) - \mathrm{dim}(\Theta_0)\) degrees of freedom (dof).

In our scenario, \(\Theta_0\) is a single point in our parameter space, i.e. \(\vec{\vartheta}_0 = (h_0, \lambda_0)\), 
while \(\Theta\) is the whole space of \((h, \lambda)\) pairs.
Moreover, all conditions of Wilk's theorem are fulfilled:
\begin{itemize}
    \item \(\Theta_0 \subset \Theta\): Since \(\Theta_0 = \{\theta_0\}\) consists of a single point \(\theta_0\) in \(\Theta = \mathbb{R} \times (0, \infty)\)
    \item The optimum parameter configuration lies in the interior of \(\Theta\): 
    \(\theta_{\mathrm{max}} = (0, \lambda)\) where \(\lambda \in (0, \infty)\) is arbitrary.
\end{itemize}
Thus, assuming sufficiently high event rates, \(\Lambda\) follows a \(\chi^2\) distribution with \(2\) dof.
As we consider counting experiments, our likelihood function is given by a product of Poisson distributions, i.e.
\begin{equation}
    \mathcal{L}(h, \lambda \; \vert \; \vec{n}) = \prod_{b = 1}^{n_f} \prod_{i=1}^n \text{Pois} (n_{ib}, \eta_{ib} (h,\lambda))\,,
    \label{eqn:NLLR2}
\end{equation}
where $\eta_{ib}$ denotes the theoretical expected value of events in bin \(i\) and flavor \(b\):
\begin{equation}
    \eta_{ib} (h,\lambda) = p_{i b}(h, \lambda) N_\text{tot} \,.
\end{equation}
Here $N_{\mathrm{tot}}$ is the total number of neutrino events.

Subsequently, the NLLR of our model reads
\begin{equation}
    \Lambda(\vec{n}) = -2 \ln{\left(\frac{\mathcal{L}(h_0, \lambda_0)}{\sup_{(h, \lambda) \in \mathbb{R}^2} \left(\mathcal{L}(h, \lambda)\right)}\right)} \,,
\end{equation}
expanding to
\begin{align}
    \Lambda(\vec{n}) = -2 \sum_{i, b} \left[ n_{ib}
    \left(\ln{(\eta_{ib}(h_0, \lambda_0))} - \ln{(\eta_{ib}(h^{\ast}, \lambda^{\ast}))}\right) 
    - \left(\eta_{ib}(h_0, \lambda_0) - \eta_{ib}(h^{\ast}, \lambda^{\ast})\right)\right]\,,
\end{align}
where \(h^{\ast}\) and \(\lambda^{\ast}\) are the parameters maximizing the likelihood on \(\Theta\).

For \(H_0\) we assume that \((h_0,\lambda_0)\), i.e. a certain combination of GW parameters, are the underlying parameters of the given data set.
We then proceed as follows:
We generate a data set representing the expected number of neutrino events if no GW is present.
Then a grid of parameter configurations \((h_0, \lambda_0)\) is scanned and the corresponding NLLR values are compared to the 
\(\alpha\)-quantile value \(\Lambda_{\alpha}\) obtained from the \(\chi^2\) distribution with \(2\) dof.
If \(\Lambda(\vec{n}) > \Lambda_{\alpha}\) the null hypothesis is rejected meaning that the GW parameters would lead to an observable effect at
\((1 - \alpha) \times 100\,\%\) confidence level.

\section{Numerical Validation of the Criteria from Section~\ref{sec:ObsCons}}
\label{app:num}
In this appendix, the criteria from section~\ref{sec:ObsCons} will be 
numerically validated through a statistical analysis. In order to estimate the actual sensitivity
of neutrino oscillations on the presence of a GW, we perform a toy analysis taking into account all 
deviations from the standard oscillation pattern. To this end, we model a hypothetical experimental 
setup and estimate the strain and frequency regions this setup would be sensitive to. It is assumed that 
this hypothetical experiment counts neutrino event rates in finite, equal width energy bins and the 
corresponding neutrino signal is expected to originate from a source at some distance $L_0$. For this
we assume a peaked neutrino energy spectrum, like for example found for neutrinos from fixed target experiments
or from decaying particles at rest, approximated by a Gaussian energy spectrum with width $\sigma_E$ centered 
at some mean energy $\bar{E}$. The number of events in each bin gets predicted analogous to appendix~\ref{app:Statistics}.
In this analysis we only consider ``plus" (+) polarized GWs perpendicular to the neutrino trajectory since 
both possible polarizations are expected to cause a similar effect. Furthermore, we assume a data taking 
period of $T_\text{exp} = 20\,\text{yr}$. For GW frequencies $f \gtrsim T_\text{exp}^{-1} \approx \SI{1.6}{\nano\hertz}$
the effect of GW induced decoherence is insensitive to the phase $\Phi^+$ of the GW since we are averaging over 
multiple cycles of the wave period. Hence, we set it to zero.

\begin{figure}
    \centering
    \includegraphics{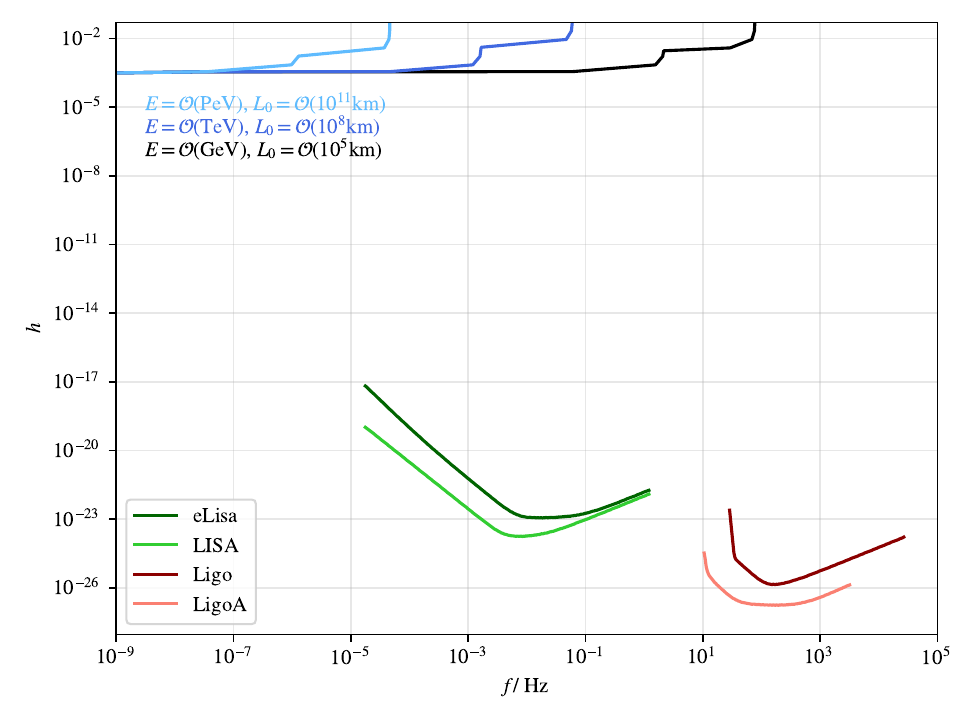}
    \caption{Upper bounds on the GW strain $h$ as a function of the frequency $f$ at 95\,\% confidence level and wave packet widths $\sigma_x = \SI{1}{\nano\meter}$ for the scenarios from table~\ref{tab:Parameters}. We also show the sensitivity curves for EPTA~\cite{Kramer:2013kea}, IPTA~\cite{Antoniadis:2022pcn}, eLISA~\cite{Amaro-Seoane:2012aqc}, LISA~\cite{Sathyaprakash:2009xs}, LIGO~\cite{LIGOScientific:2010ped}, LIGOa~\cite{LIGOScientific:2010ped} and Virgo~\cite{LIGOScientific:2010ped}.}
    \label{fig:Std_best}
\end{figure}

To validate the criteria from section~\ref{sec:ObsCons}, we now consider the scenario of either free 
falling artificial neutrino sources placed, for example, in a stable Lagrange point within the solar system 
or neutrinos originating from dark matter decaying or annihilating inside the gravitational wells of celestial 
bodies like the moon. Therefore, we assume an initially pure muon neutrino flux modelled by a Gaussian shaped energy 
spectrum with a total event count of $N = 10^5$. We select a 
neutrino wave packet width in the $\si{\nano\meter}$ range~\cite{DayaBay:2016ouy},
and conduct a statistical analysis using the parameters in table~\ref{tab:Parameters} to verify the 
estimations from section~\ref{sec:ObsCons}. For details of the statistical analysis see appendix~\ref{app:Statistics}.
This enables us to determine the strains and frequencies
of GWs to which corresponding experiments would be sensitive to.
\begin{table}
    \caption{Values of the chosen
    baselines $L_0$, various energies $E$ and bin 
    widths $\Delta E$, where effects would be observable 
    for SM neutrino masses and wave packet widths 
    $\sigma_x = \SI{1}{\nano\metre}$. The oscillation lengths 
    for the different energies are given in table~\ref{tab:uncertainty_of_source_location}.}
    \begin{center}
    \begin{tabular}{c c c c} 
    \toprule
    Scenario  & $L_0 \, / \, \si{\kilo\meter}$ & $E$ & $\Delta E$ \\
    \midrule
    1 & $\thicksim \num{e5}$ & $[1, 10] \si{\giga\electronvolt}$ & $\thicksim \SI{0.1}{\giga\electronvolt}$\\ 
    2 & $\thicksim \num{e8}$ & $[1, 10] \si{\tera\electronvolt}$ & $\thicksim \SI{0.1}{\tera\electronvolt}$\\
    3 & $\thicksim \num{e11}$ & $[1, 10] \si{\peta\electronvolt}$ & $\thicksim \SI{0.1}{\peta\electronvolt}$\\
    \bottomrule
    \end{tabular}
    \label{tab:Parameters}
    \end{center}
\end{table}
Figure~\ref{fig:Std_best} shows the upper limit of the strain $h$ as a function of the frequency $f$ at $95\,\%$ CL 
assuming the absence of the GW induced decoherence in the experimental data set for the parameters from table~\ref{tab:Parameters}
with sensitivity curves for the experiments taken out of reference~\cite{Moore:2014lga}.
LIGO and LIGOa give limits at 90\,\% CL~\cite{LIGOScientific:2010ped} as well as Virgo~\cite{LIGOScientific:2010ped},
and LISA~\cite{Sathyaprakash:2009xs}. EPTA, IPTA, and eLISA give limits at 95\,\% CL~\cite{Kramer:2013kea, Antoniadis:2022pcn, Amaro-Seoane:2012aqc}. 

Figure~\ref{fig:Std_best} demonstrates that the criteria established in section~\ref{sec:ObsCons} provide 
accurate estimations. Moreover, it becomes apparent that a neutrino oscillation experiment lacks the sensitivity
required to probe realistic parameters of coherent GWs. Therefore, it is more promising to investigate the 
influence of the SGWB on neutrino oscillations, as explored in section~\ref{sec:sgw}.

\end{document}